\RequirePackage{fix-cm}
\documentclass[11pt,oneside]{article}  
\usepackage[left=2cm, right=2cm, top=2cm]{geometry}
\usepackage{subfigure}
\usepackage{float}
\usepackage[numbers]{natbib}
\usepackage{graphicx}

\usepackage{hyperref} 
\hypersetup{
	colorlinks=true,	
	linkcolor=blue,		
	citecolor=blue,		
	filecolor=blue,		
	urlcolor=blue		
}

\begin{document}

 \title{Real-time Broadband RFI Excision for the Upgraded GMRT}   




\author{Ruta Kale, Kaushal D. Buch, Sanjay Kudale, Mekhala Muley, Ajith Kumar B.}
\maketitle
\begin{center}
    NCRA Technical Report No. R1401, March 4, 2020 \\
    National Centre for Radio Astrophysics, TIFR, Pune, India
\end{center} 
\tableofcontents




\section{Introduction}
 Radio Frequency Interference (RFI) mitigation is a major challenge for all the radio observatories across the world. It adversely affects the radio observations, making it difficult to achieve the theoretical sensitivity and can create signal distortion if not mitigated early in the processing chain. Over the years, a wide range of mitigation techniques have been proposed at different locations in the receiver system of a radio telescope \cite{ford2014rfi, baan2019implementing}. 

Two broad strategies to mitigate RFI are: i) offline method: 
after the data (visibilities or beam) has been recorded, and ii) online method: 
on-the-fly at the highest possible time resolution before it is processed or recorded in the final form 
accessible to the user. In the online mitigation, one does not record the ``raw'' form of the data 
on which the mitigation happens, and thus there is permanent loss of the samples that are selected for 
excision. On the other hand, offline methods are inadequate 
as they generally operate on data recorded at a time resolution much coarser than Nyquist sampling. 
This limits their capability to excise the RFI as compared to the online methods. The online methods can do much damage control with a lower amount of flagging.

A real-time broadband RFI mitigation system has been developed for the Upgraded GMRT (uGMRT) \cite{buch2016towards,buch2018implementing}. Here, we report 
the engineering, imaging, and beam/pulsar results from the test observations in all the bands of the uGMRT during the monsoon of 2018 and 2019. Broadband power-line RFI is almost always present in uGMRT bands 2 (120 - 240 MHz), 3 (300 - 500 MHz) and 4 (550 - 850 MHz), but the monsoon period is chosen to understand the performance in the worst RFI situations arising due to an increase in the wind-speed and moisture.

\subsection{RFI at the uGMRT}
\begin{figure}[h]
\begin{center}
 \includegraphics[trim={12.7cm 10cm 0cm 0cm}, clip,height=10cm]{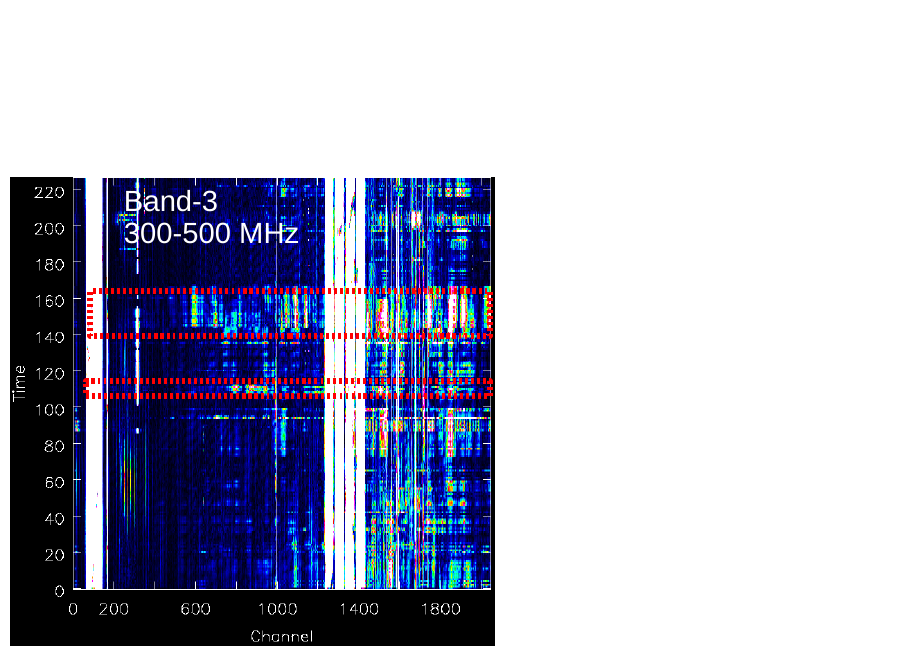}
 \advance\leftskip-3cm
\advance\rightskip-3cm
 \caption{The time-frequency plane data from the uGMRT at band 3 for a baseline between central square antennas is shown. A couple of examples of BB RFI are marked by the red dashed boxes. The lowest channel number is at the highest frequency.}
 \label{allbandrfi}
 \end{center}
\end{figure}

In terms of its frequency-domain characteristics, RFI can be categorized as Broadband (BB) or Narrowband (NB). In the time-domain, BB RFI manifests as short duration, impulsive bursts of interference signal, typically from spark-like events (e.g. lightning discharge, power line discharge, spark ignition discharge); whereas NB RFI is due to spectrally-confined sources such as transmissions from communication systems - terrestrial or otherwise. Example of uGMRT band-3 data affected by NB and BB RFI is shown in Fig.~\ref {allbandrfi}. The broadband RFI is highlighted with boxes.

BB RFI is stronger at low frequencies (below 1 GHz) and hence its excision is important for observing in bands 2, 3 and 4 of the uGMRT. Details of power-line RFI at GMRT and the need for mitigation from an earlier study are described in the reports \cite{rfirep01,rfirep02}. Of particular interest to this work is that these reports describe the power-line RFI and its effect on the GMRT observations, along with various measurements carried out in and around the array. It also describes the harmful levels of interference, effect on the sensitivity due to power-line RFI, with an emphasis on the need for RFI mitigation.

\subsection{Need for real-time excision}

\begin{enumerate}
\item For temporally impulsive RFI (Fig.~ \ref{rfipl}), the total energy spreads in the Fourier domain and hence excision is needed before the Fast Fourier Transform (FFT) stage. Power-line RFI has a low duty cycle but a high spectral occupancy, and hence, a frequency selective filter (like a notch or band-stop) does not work in this case. 

\begin{figure}[H]
\begin{center}
 \includegraphics[scale=0.7]{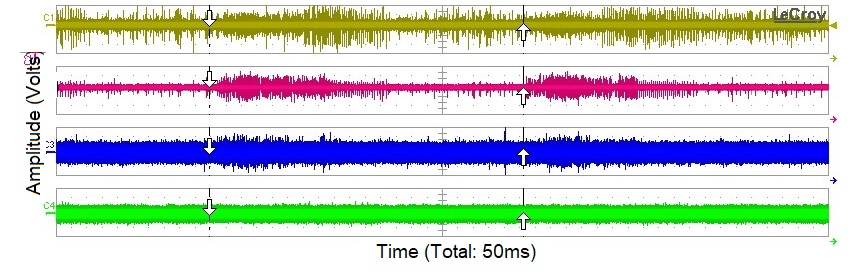}
 \advance\leftskip-3cm
\advance\rightskip-3cm
 \caption{A typical example of uGMRT band-2 (C01 antenna, first subplot), band-3 (C02 antenna, second subplot), band-4 (C05 antenna, third subplot) and band-5 (C08 antenna, fourth subplot) data acquired simultaneously at the input of GWB (100 MHz bandwidth) using a 4-channel mixed-signal oscilloscope with 5ns sampling; 50ms data shown in the plot (X-axis: Time (20.33ms between the dotted markers),Y-axis: Voltage, 500mV/division for band-2 and band-3 100mV/division for band-4 and 1V/division for band-5). The periodically repeating impulses are from the power-line interference.}
 \label{rfipl}
 \end{center}
\end{figure}

\item RFI is correlated in closely-spaced antennas and hence excision is best suited in the pre-correlation domain. This is especially important as the GMRT array has closely spaced antennas in the central square.


\end{enumerate}

\section{Online BB RFI Mitigation for the uGMRT}
BB RFI is dominant at lower frequencies ($<$1GHz) and occurs due to gap discharge or corona discharge on high-power transmission lines and associated transmission equipment in the vicinity of the GMRT array. The online (real-time) BB RFI filter for the GMRT Wideband Backend (GWB) operates on digitized (8-bit signed integer) time-domain Nyquist-sampled signal on Field Programmable Gate Array (FPGA)  \cite{buch2019real}. The statistical estimation of the signal is carried out using Median-of-MAD (MoM) estimator. There are two detection techniques available: (a) Voltage detection and (b) Power detection. The filtering  (replacement) options available are digital noise, constant value and threshold. In this report, the voltage detection and the results using this technique are described.

 Median Absolute Deviation (MAD) is computed on Nyquist-sampled digitized time series for each antenna and polarization ($MAD = median (abs(X_i - median(X)))$) on 16k data samples.
 To get a more robust estimation, the median of 16k MAD
values are computed for getting the Median-of-MAD
(MoM).  To meet the real-time requirements, this estimation is carried out on every fourth input sample out of the samples arriving at a 1.25ns interval. (Calculation: $(16384 samples*16384 MAD values*1.25*10^{-9} sampling period)*4 = 1.34s$). Thus, 1.34s worth of data goes into the real-time statistics for the RFI filtering
and is applied to the subsequent data.
Values outside the threshold 
$median \pm n*1.4826*MoM$ are treated as RFI and are
replaced either by a constant value or digital noise, or the
threshold value. $n$ is kept as $3$ for a $3\sigma$ threshold.

RFI counter keeps a record of the number of detected
RFI samples and the total number of samples per antenna and
polarization. The ratio of these counts would provide the fraction of flagged samples over user-defined time intervals (minimum duration of 5 minutes as per current configuration) during the observation. Example of processed data from the counter output is shown in Section \ref{counterpercent}.\\

\subsection{Testing scheme for the real-time RFI filter: System setup and configuration}

\begin{figure}[h]
\begin{center}
 \includegraphics[scale=0.3]{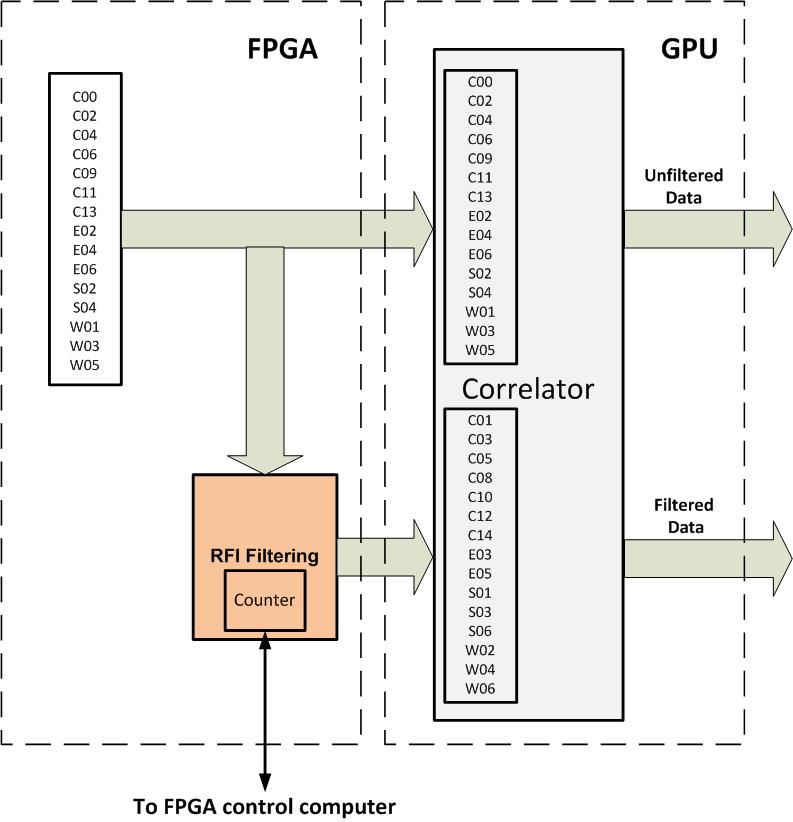}
 \centering
 \caption{GMRT antenna configuration in the 1:2 digital copy mode. In this mode 
signals from half the antennas are copied into the rest half. One set  
passes through the filter, and the unfiltered copy is obtained using the signals 
from the other set of antennas.}
 \label{rfidigcpy}
 \end{center}
\end{figure}

\begin{table}[h]
\caption{The set up in 1:2 mode.} \label{anttable}
\vspace{0.2cm}
\resizebox{\textwidth}{!}{%
\begin{tabular}{|c|c|c|c|c|c|c|c|c|c|c|c|c|c|c|c|c|c|c|c|c|}
\hline
 Unfiltered& C00& C02 &C04 &C06 &C09 &C11 &C13 &E02 &E04 &E06 &S02 &S04 &W01 &W03 &W05\\
AIPS no.&   01 &  03  &05  &07  &09  &11  &13  &15  &17  &19  &21  &23  &25  &27  &29 \\
\hline
Filtered& C01 &   C03 &C05 &C08 &C10 &C12 &C14 &E03 &E05 &S01 &S03 &S06 &W02 &W04 &W06\\
AIPS no.&   02&   04  &06  &08  &10  &12  &14 & 16  &18  &20  &22  &24  &26  &28  &30  \\
\hline
\end{tabular}}
\label{annttable}
\end{table}
The quality of RFI filtering was tested through a comparative analysis between unfiltered and filtered signals of the same antenna for the same instance of time. This scheme was implemented by configuring the antenna input and the corresponding GWB parameters to support simultaneous processing for the same antenna signal through two chains – one with the RFI filter and one without.

Data for imaging tests were taken with the mode of $digital copy$. In this mode, signals from half the antennas were digitally copied (inside the FPGA) into the rest half. One set was 
passed through the filter, and the unfiltered copy was obtained in the signals 
from the other set of antennas. The schematic of this is shown in  Fig.~\ref{rfidigcpy}.
The mode is referred to as a $1:2$ digital copy mode. The mode where the online RFI filter was used 
on all the antennas without retaining a copy of the unfiltered signal is referred to as $1:1$. For each observation the counter values were recorded to provide an estimate of the number of samples affected by the RFI excision. All the test described in this report have been carried out using the 1:2 digital copy mode.

For the imaging tests, the filtered and unfiltered data were recorded in the same long-term accumulation (.lta) file and then converted to FITS format using the standard $listscan$ and $gvfits$ programs. The names of antennas that were filtered and unfiltered were provided in the log file produced by $listscan$, and separate FITS files were created for further analysis. 
Table~\ref{annttable} provides antenna names for which the filtered and unfiltered copy 
of the signal is present. The AIPS antenna IDs are also given for reference.
Data analysis was carried out using the NRAO Astronomical Image Processing System (AIPS) 
and Common Astronomy Software Applications (CASA).

\newpage
\section{Results from engineering tests}\label{engtest}

We carried out tests in the monsoon of 2018 and 2019, primarily in the uGMRT bands-3 and 4 over an RF bandwidth of 200 MHz to understand the effect of RFI filtering on the beam and interferometer outputs.


 %

An example of an engineering test carried out on 26 July 2018 in Band-3 is presented. The test was carried out using the 1:2 filtering mode on 3C147 and a processing bandwidth of 200 MHz. Beam and interferometer data were recorded simultaneously at a time-resolution of 81.92$\mu s$ and 671ms, respectively. The plots are for a single spectral channel corresponding to 451 MHz. This spectral channel has an expected power level and is away from the channels affected by NB RFI. The cross-correlation function, closure phase, beam and beam-FFT, and percentage flagging are described. 

\subsection{Effect on closure phase}\label{closure}

\begin{figure}[h]
 \hbox{\hspace{-2cm}\includegraphics[width=1.2\textwidth]{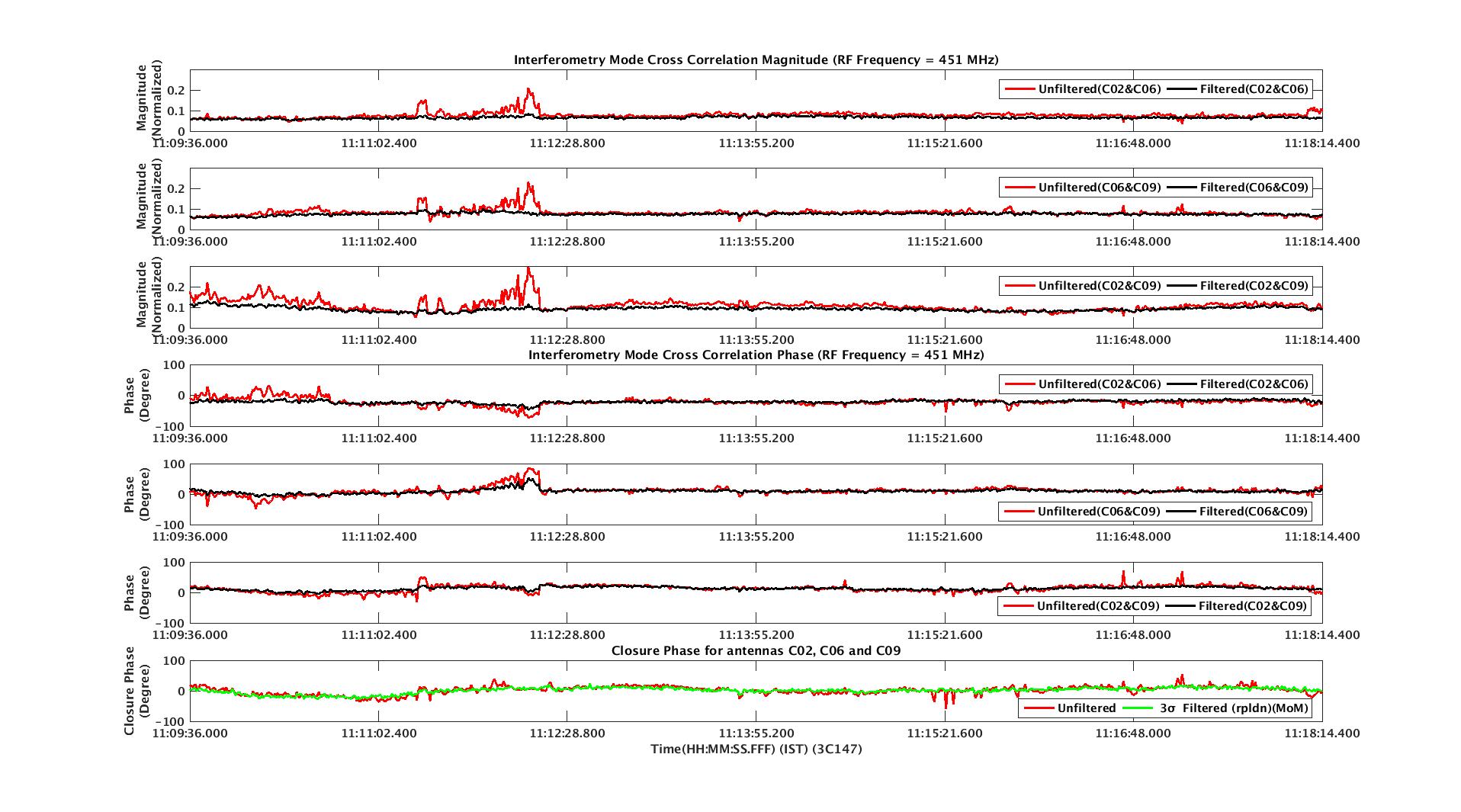}}
 \centering
 \vspace{-1cm}
 \caption{Closure phase comparison between antennas C02,C06 and C09. The closure phase computed from the data in subplots 4,5, and 6 is shown in subplot 7. In this case, RFI causes error in the closure phase up to $\pm 60 $ degrees, which is reduced to zero(within $
 \pm 5 $ degrees) in the filtered data. [Sec.~\ref{closure}]}
 \label{rficp}
\end{figure}

 The filtered and unfiltered copy of a single spectral channel time series for a 3$\sigma$ threshold and digital noise replacement is shown. Fig.~ \ref{rficp} shows the closure phase between three closely spaced antennas of the central square (C02, C06, and C09) for unfiltered and filtered outputs. The first three subplots show the cross-correlation amplitude for three baselines. Subplots 4,5, and 6 show the phase of the cross-correlation. The closure phase computed from the data in subplots 4,5, and 6 is shown in subplot 7. RFI causes fluctuations in the closure phase up to about $\ pm60$ degrees, which is brought close to zero (within $\ pm5$ degrees) in the filtered data.
 \subsection{Effect on the Cross-correlation function (CCF)} \label{ccf}
 
\begin{figure}[h]
 \begin{center}
 \includegraphics[scale=0.2]{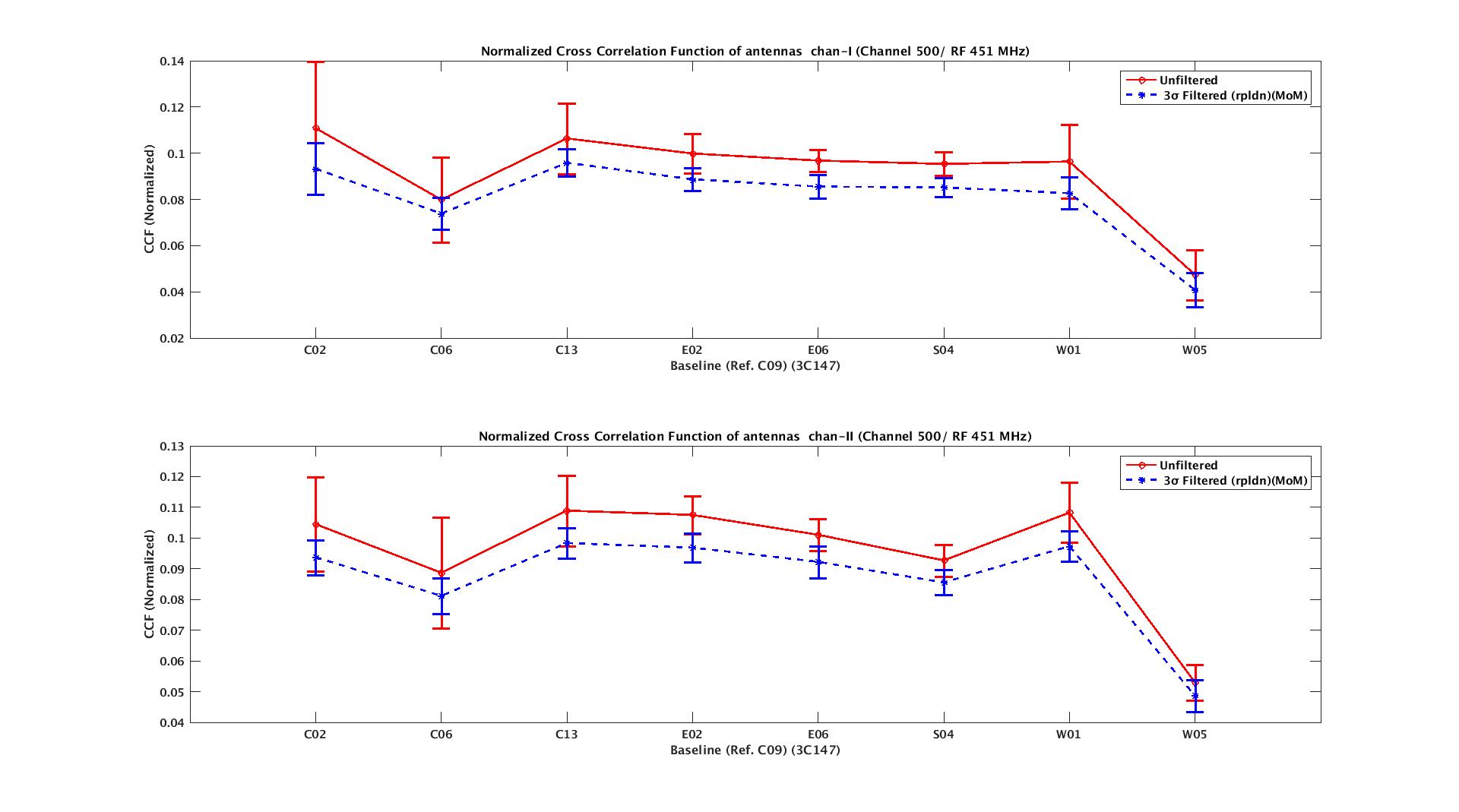}
 \centering
 \caption{CCF comparison between antennas with reference to C09 antenna. The ratio of the standard deviation of unfiltered to filtered data is in the range of 1.5 to 3. It can be seen that where there was no BB RFI, the error bar has been unaltered by the filtered output (see C09-E06 baseline for channel-II). [Sec.~\ref{ccf}]}
 \label{rficcf}
 \end{center}
\end{figure}

 Presence of BB RFI causes larger fluctuations in the measurement of the cross-correlation function. This is particularly prominent on shorter baselines (The central square and a few arm antennas) where the RFI between the antennas is correlated. Fig. \ref{rficcf}  shows that the ratio of the standard deviation of unfiltered to filtered data is in the range of 1.5 to 3. It can be seen that where there was no BB RFI, the error bar has been unaltered by the filtered output (see C09-E06 baseline for channel-II in Fig.~\ref {rficcf}).
 
 \subsection{Simultaneous beam and interferometer comparison}\label{beamint}
\begin{figure}[h]
 \begin{center}
 \includegraphics[scale=0.3]{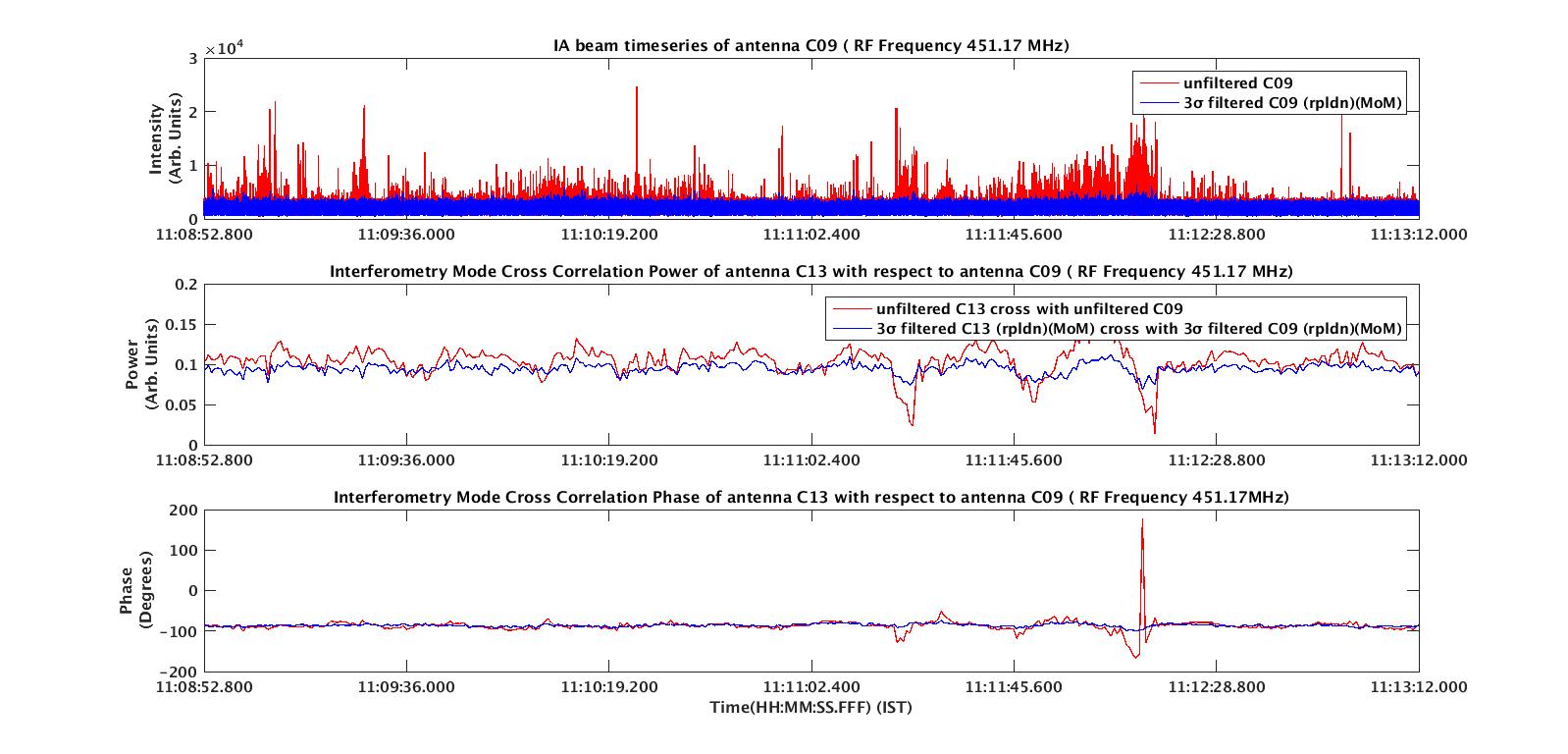}
 \centering
 \caption{Comparison between beam and interferometry simultaneous observation. The simultaneous cross-correlation amplitude (subplot 2) and phase (subplot 3) show that both quantities remain stable while their unfiltered counterparts fluctuate during the event of BB RFI. A maximum improvement of 7.7 dB (ratio of unfiltered to filtered value) is observed in the beam data shown in subplot 1. [Sec.~\ref{beamint}]}
 \label{rfibi}
 \end{center}
\end{figure}

 Time-series of single spectral channel show improvement between the unfiltered and filtered beam data in subplot 1 of Fig.~\ref {rfibi}. The simultaneous cross-correlation amplitude (subplot 2) and phase (subplot 3) show that both quantities remain stable while their unfiltered counterparts fluctuate during the event of BB RFI. A maximum improvement of 7.7 dB (ratio of unfiltered to filtered value) is observed in the beam data shown in subplot 1.

 \subsection{Beam data analysis}\label{beamdata}
 
\begin{figure}[h]
\begin{center}
 \includegraphics[scale=0.3]{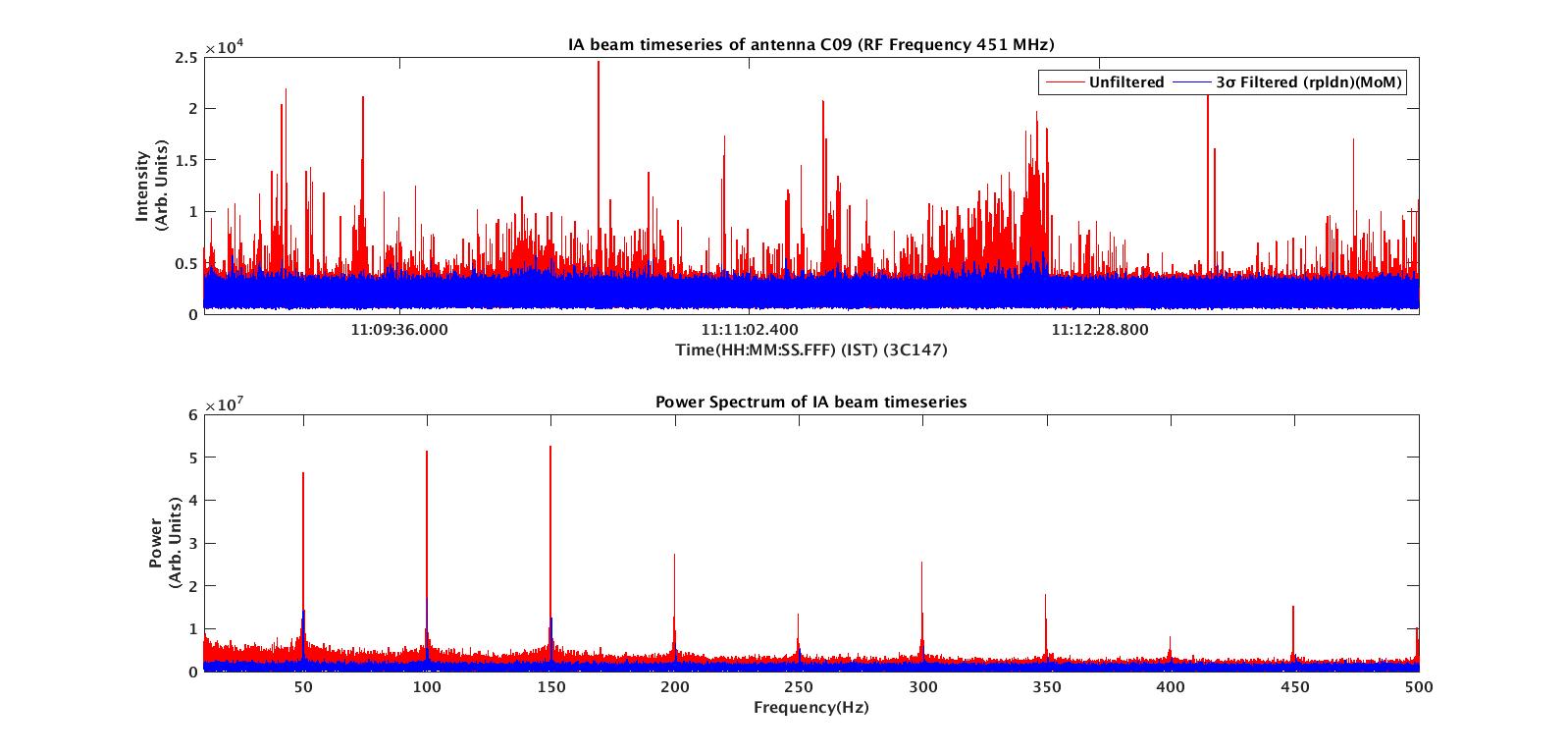}
 \centering
 \caption{Single spectral channel beamformer output and its Fourier transform. The 100 Hz, 200 Hz and harmonics are seen in the unfiltered data, which are reduced by a factor of two (100 Hz) and four (200 Hz) in the filtered data set. These harmonics correspond to the multiples of power-line frequency (50 Hz). [Sec.~\ref{beamdata}]}
 \label{rfibmfft}
 \end{center}
\end{figure}

 Time series of single spectral channel beam data is shown in subplot 1 of Fig.~\ref {rfibmfft}. The second subplot shows the Fourier transform of this time series. The 50 Hz, 100 Hz, and harmonics are seen in the unfiltered data, which are significantly removed in the filtered data set. These harmonics correspond to the multiples of the 50 Hz power-line frequency. The power in the 50 Hz and 100 Hz frequencies goes down by a factor of four for the filtered data.
 
 \subsection{Antenna-based percentage flagging} \label{antflag}
\begin{figure}[h]
 \begin{center}
 \includegraphics[scale=0.65]{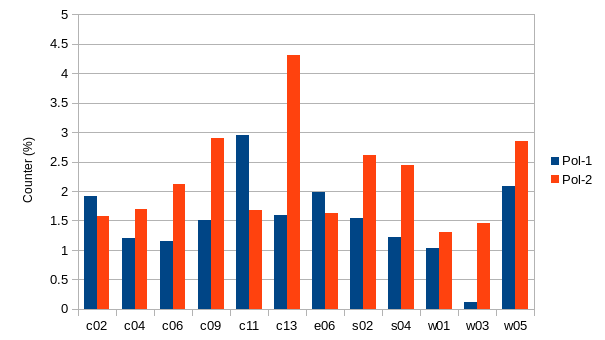}
 \centering
 
 \caption{Bar chart showing the percentage fraction of RFI samples detected and replaced during the observation. In this case, the average RFI excised is less than 2\%. [Sec.~\ref{antflag}]}
 \label{rficnt}
 \end{center}
\end{figure}

 The total fraction (in percentage) of samples replaced during this test for each antenna and polarization is shown in Fig.~\ref {rficnt}. This is generated from the counters within the FPGA design. For this test observation, the average RFI excised is less than 2\%. 
 It is noted that the flagging percentage can differ significantly between the two polarizations of the same antenna.
 
\clearpage

\newpage
\section{Beamformer data analysis and Pulsar observations}

\subsection{3C48}
As part of the engineering tests, 3C48 was observed at Band-4 on 14 June 2018. The observations were carried out in the 1:2 digital copy mode with voltage technique ($3 \sigma$ threshold). The effect of BB RFI filtering was observed using the GMRT Pulsar Tool ($gptool$).

\begin{figure}[h]
\minipage{0.45\linewidth}
 \includegraphics[scale =0.245]{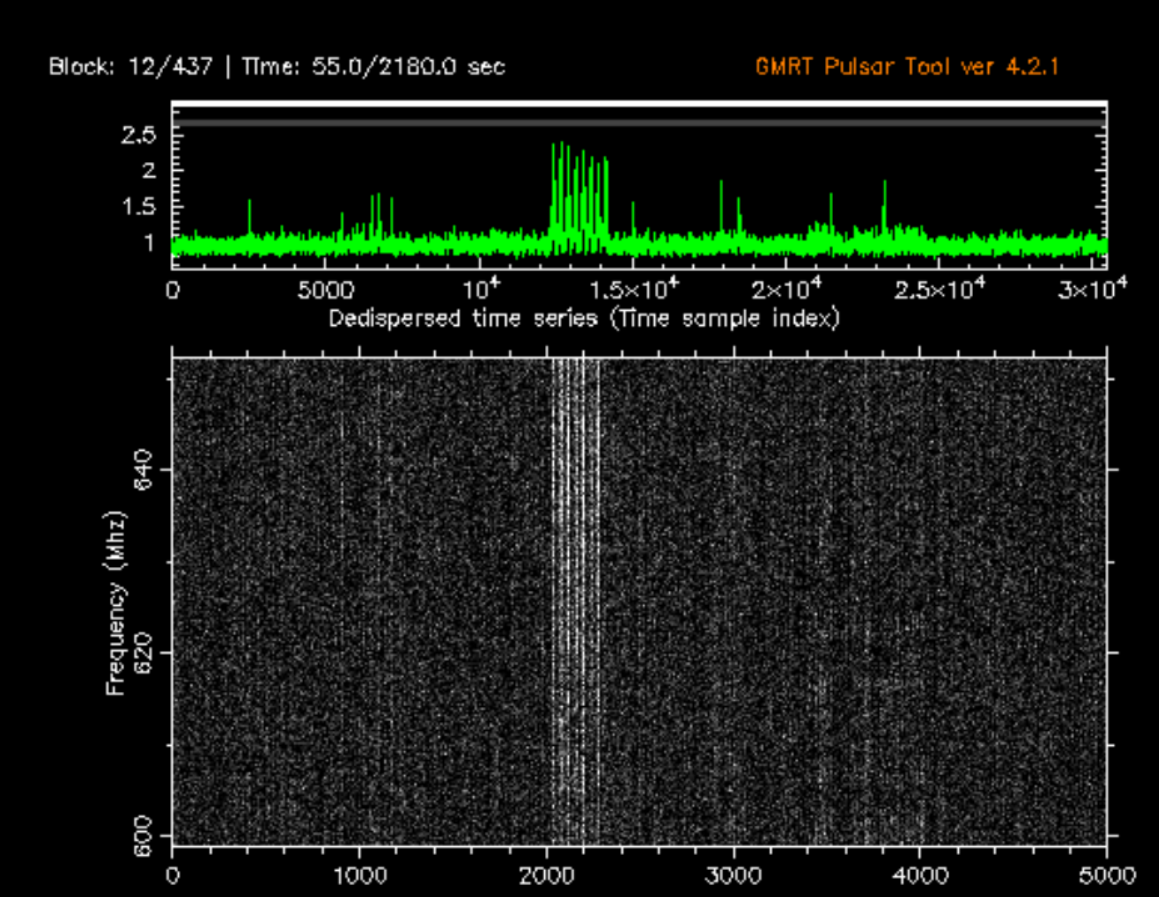}
 \centering

 \caption{Beamformer mode for 3C48(unfiltered copy): observation carried out in the digital copy (1:2) mode.}
 \label{rfibeamuf1}
\endminipage
 \hspace{0.05\linewidth}
\minipage{0.45\linewidth}
 \includegraphics[scale =0.3]{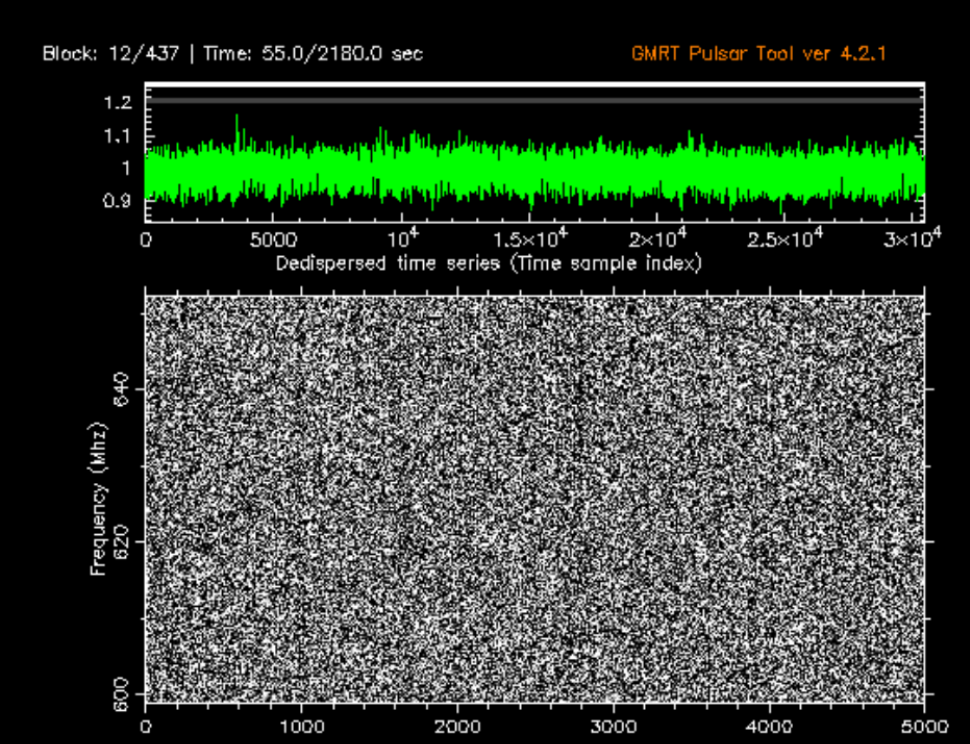}
 \centering
 \caption{Beamformer mode for 3C48 (filtered copy): observation carried out in the digital copy (1:2) mode.}
 \label{rfibeamf1}
\endminipage

\minipage{0.45\linewidth}
 \includegraphics[scale =0.3]{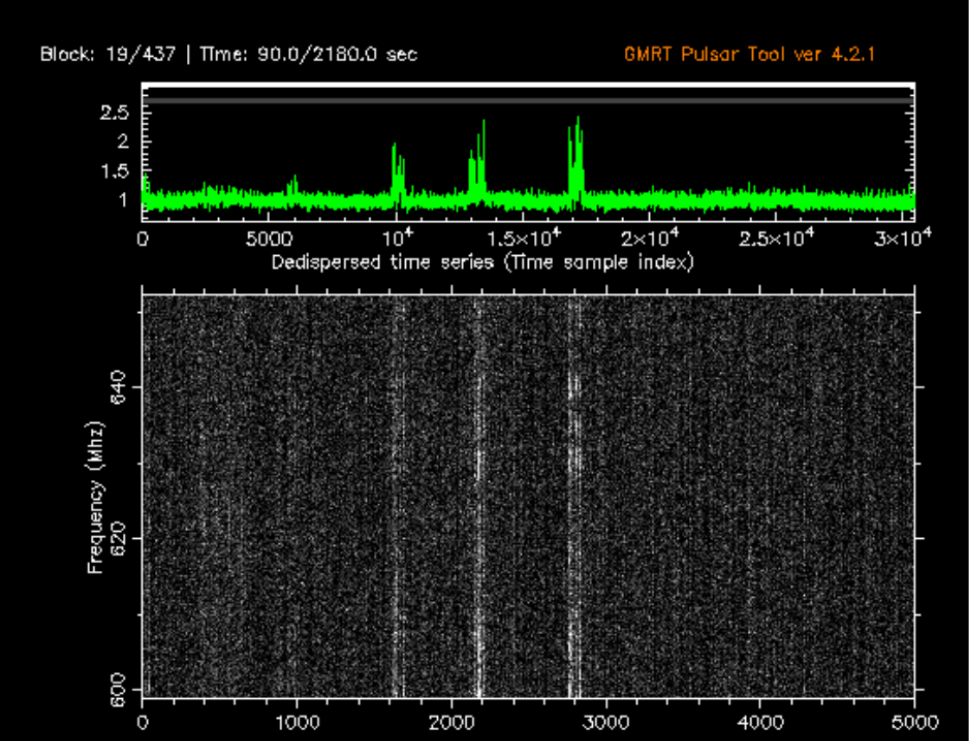}
 \centering
 \caption{Beamformer mode for 3C48(unfiltered copy): observation carried out in the digital copy (1:2) mode.}
 \label{rfibeamuf2}
\endminipage
 \hspace{0.05\linewidth}
\minipage{0.45\linewidth}
 \includegraphics[scale =0.3]{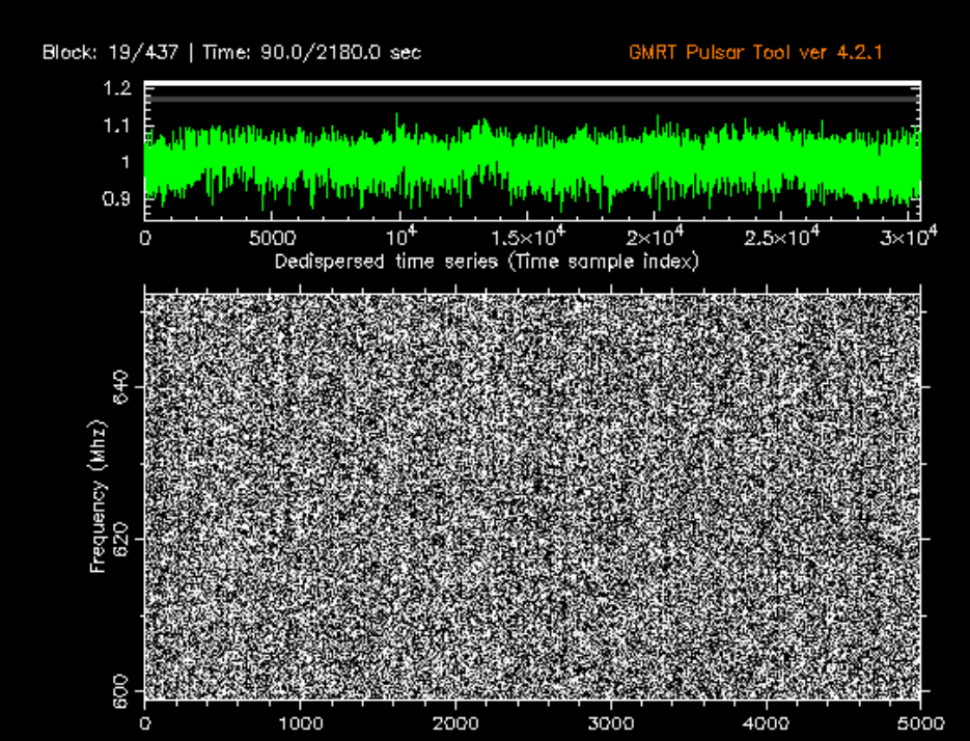}
 \centering
 \caption{Beamformer mode for 3C48 (filtered copy): observation carried out in the digital copy (1:2) mode.}
 \label{rfibeamf2}
\endminipage
\end{figure}

As an example, two blocks from the observation of 3C48 are shown with the unfiltered and filtered counterparts (Fig.~ \ref{rfibeamuf1}, ~ \ref{rfibeamuf2} and Fig.~ \ref{rfibeamf1}, ~ \ref{rfibeamf2}). In order to isolate and find out the effect of real-time BB filtering, the RFI filtering within the $gptool$ was disabled, and the processing was done with zero Dispersion Measure (DM) settings. It can be seen that the bursts of interference (which are broadband in nature) have been completely removed in the filtered data.





\subsection{PSR B0329+54 and PSR J0418-4154}

Pulsar observations were carried out in the incoherent array (IA) mode on 5th July 2019 at Band-3. 
Two pulsars, B0329+54 and J0418-4154, were observed during this session. The purpose
of this test was to check the effect of filtering on strong single pulses and averaged folded profiles.
The observation was carried out with 14 antennas in the 1:2 digital copy mode and with RFI filtering in the voltage
detection mode at 3-sigma threshold and replacement by digital noise. The bandwidth was 200 MHz in the 300-500 MHz band with 4096 spectral channels and a time-resolution of 327.68 $\mu s$.

\begin{figure}[h]
\minipage{0.45\linewidth}
 \includegraphics[scale =0.345]{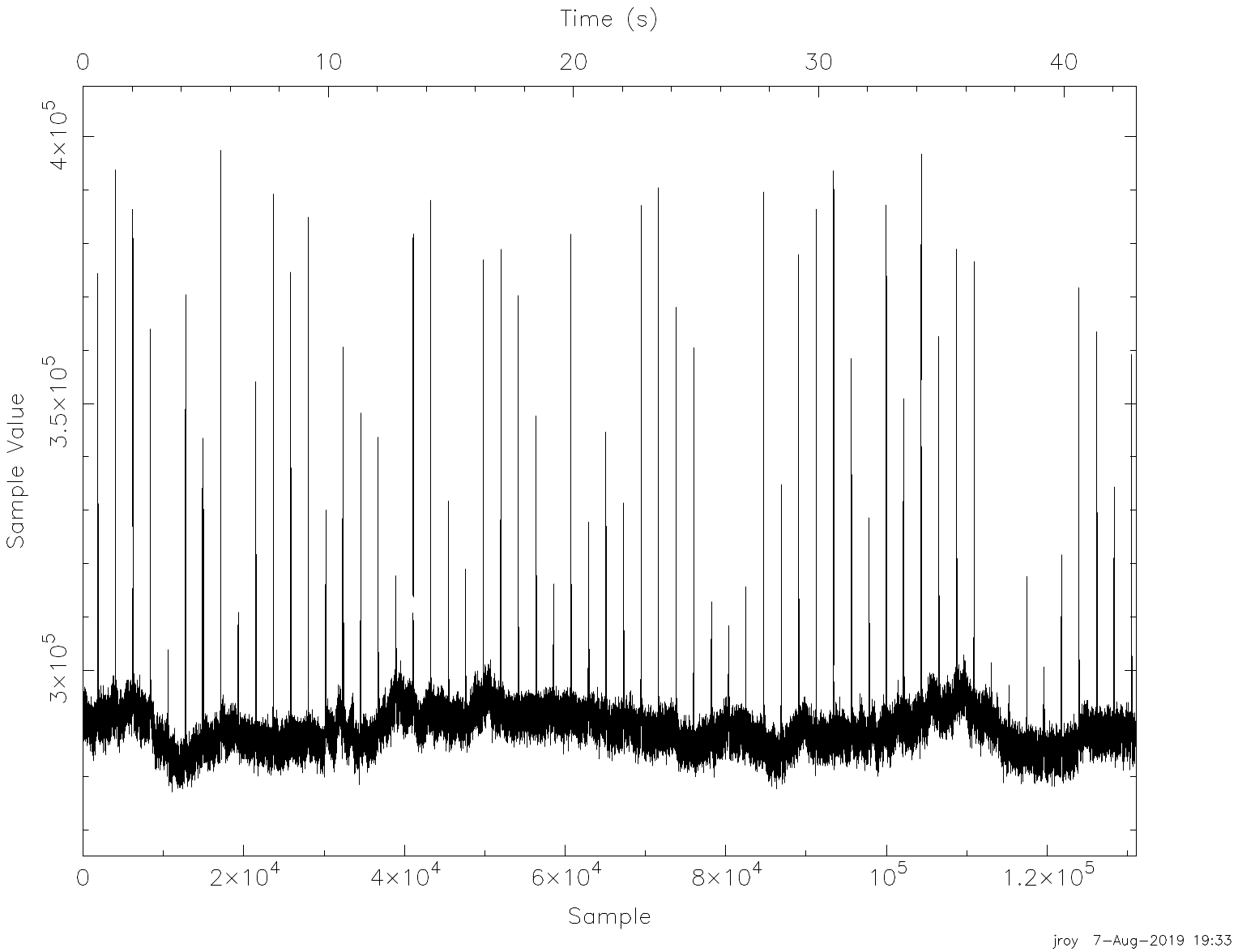}
 \centering

 \caption{Unfiltered de-dispersed time-series from a strong pulsar B0329+54 : observation carried out in the digital copy (1:2) mode.}
 \label{psrtimeu}
 \endminipage
\hspace{0.05\linewidth}
\minipage{0.45\linewidth}
\includegraphics[scale=0.345]{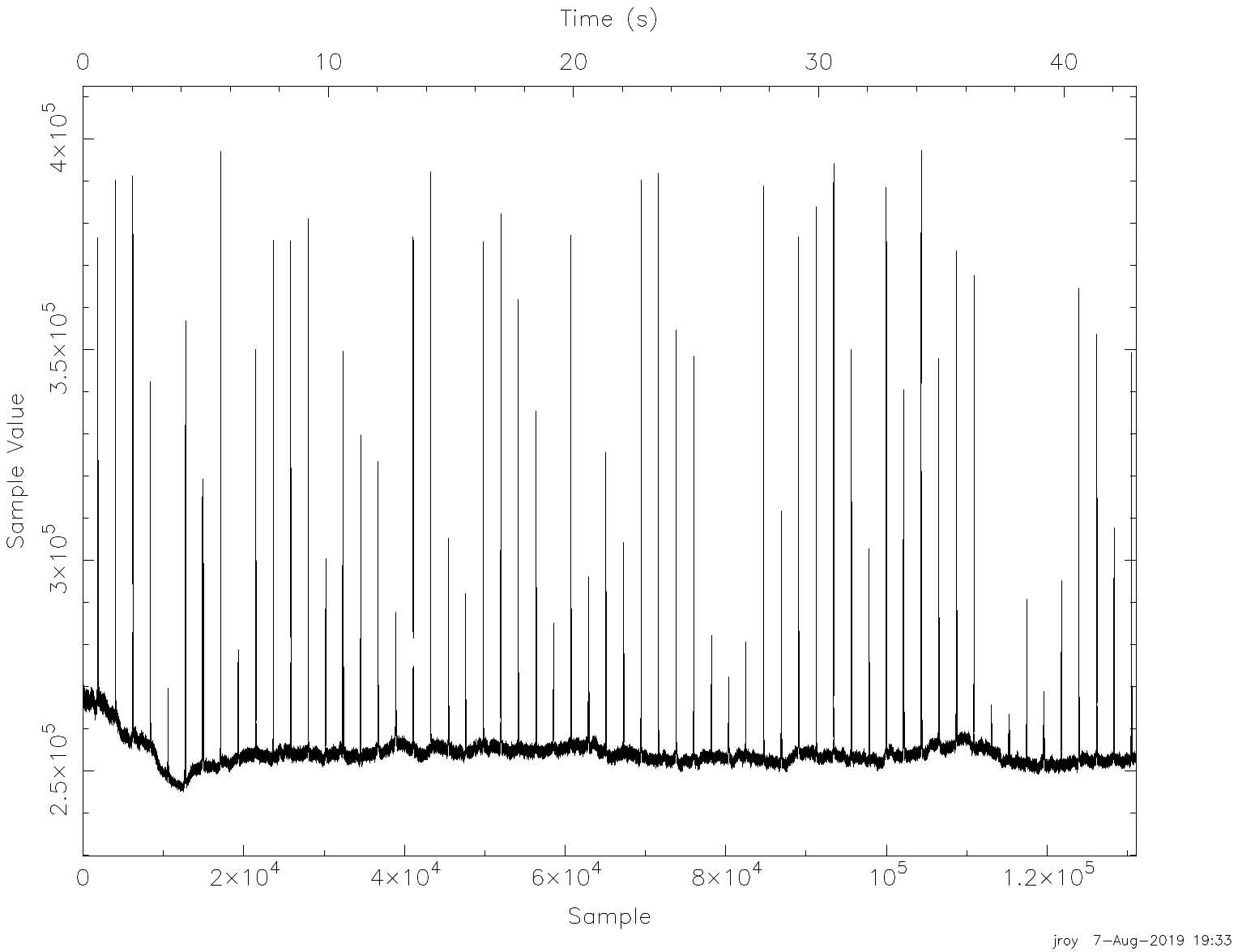}

 \centering
 \caption{Filtered de-dispersed time-series from a strong pulsar B0329+54 : observation carried out in the digital copy (1:2) mode. SNR improvement by a factor of 4.}
 \label{psrtimef}
 \endminipage
\end{figure}

The de-dispersed time-series from a strong pulsar B0329+54 shows an improvement of SNR by a factor
of 4 (Fig. \ref{psrtimef}) over the unfiltered version (Fig. \ref{psrtimeu}). The SNR of the unfiltered is 60, which increases to 240 after filtering.

\begin{figure}[H]
\minipage{0.45\linewidth}
 \includegraphics[scale =0.345]{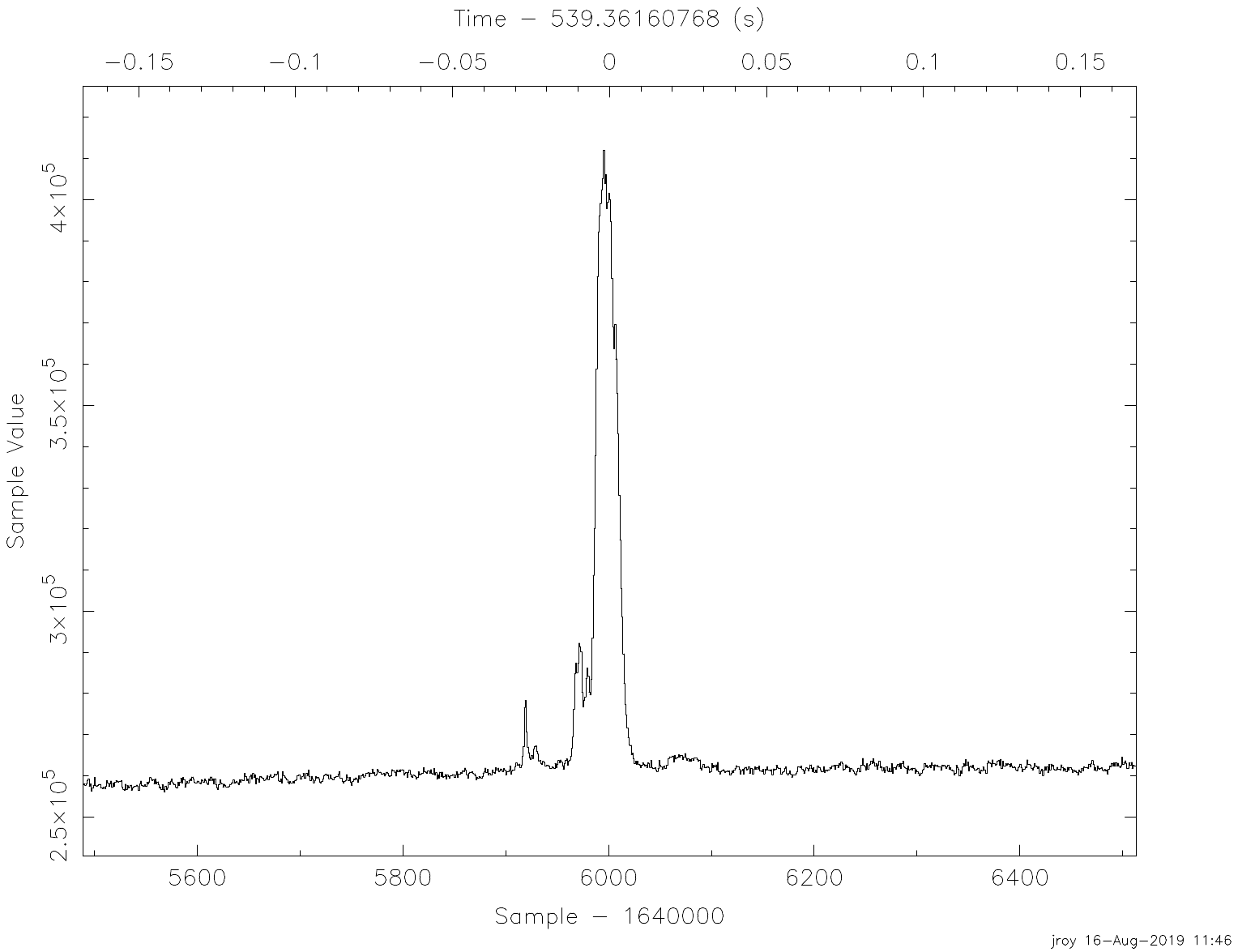}
 \centering

 \caption{Single pulse (unfiltered) from a strong pulsar B0329+54}
 \label{prsclipu}
 \endminipage
\hspace{0.05\linewidth}
\minipage{0.45\linewidth}
\includegraphics[scale=0.345]{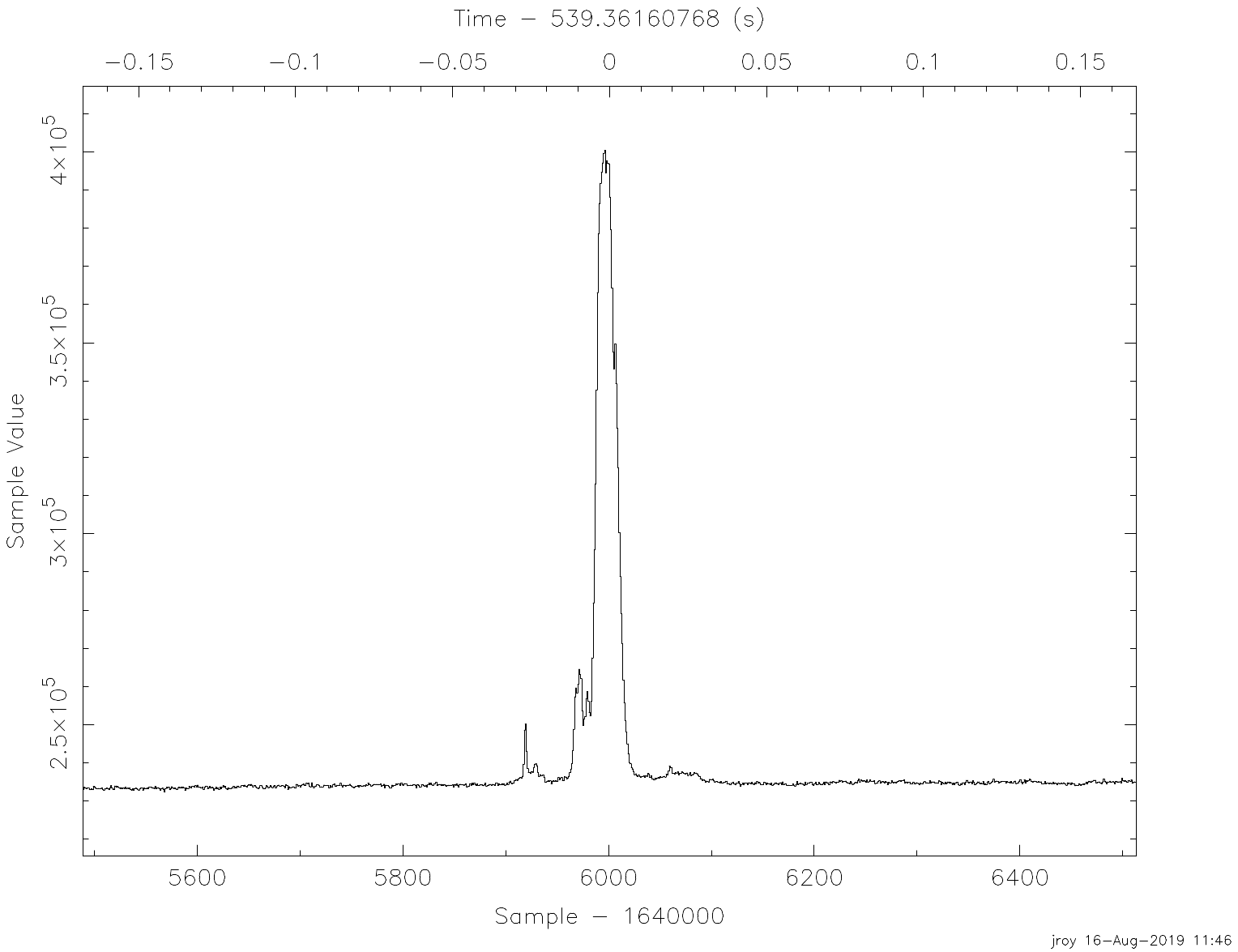}

 \centering
 \caption{Single pulse (filtered) from a strong pulsar B0329+54.}
 \label{prsclipf}
 \endminipage
\end{figure}

The strong single pulse from
PSR B0329+54 seems not to have clipped during filtering (Fig. \ref{prsclipu} and Fig. \ref{prsclipf}). These are initial results, and a detailed analysis will be carried out to quantify whether there is any change in profile shape by filtering.

\begin{figure}[h]
\minipage{0.45\linewidth}
 \includegraphics[scale =0.345]{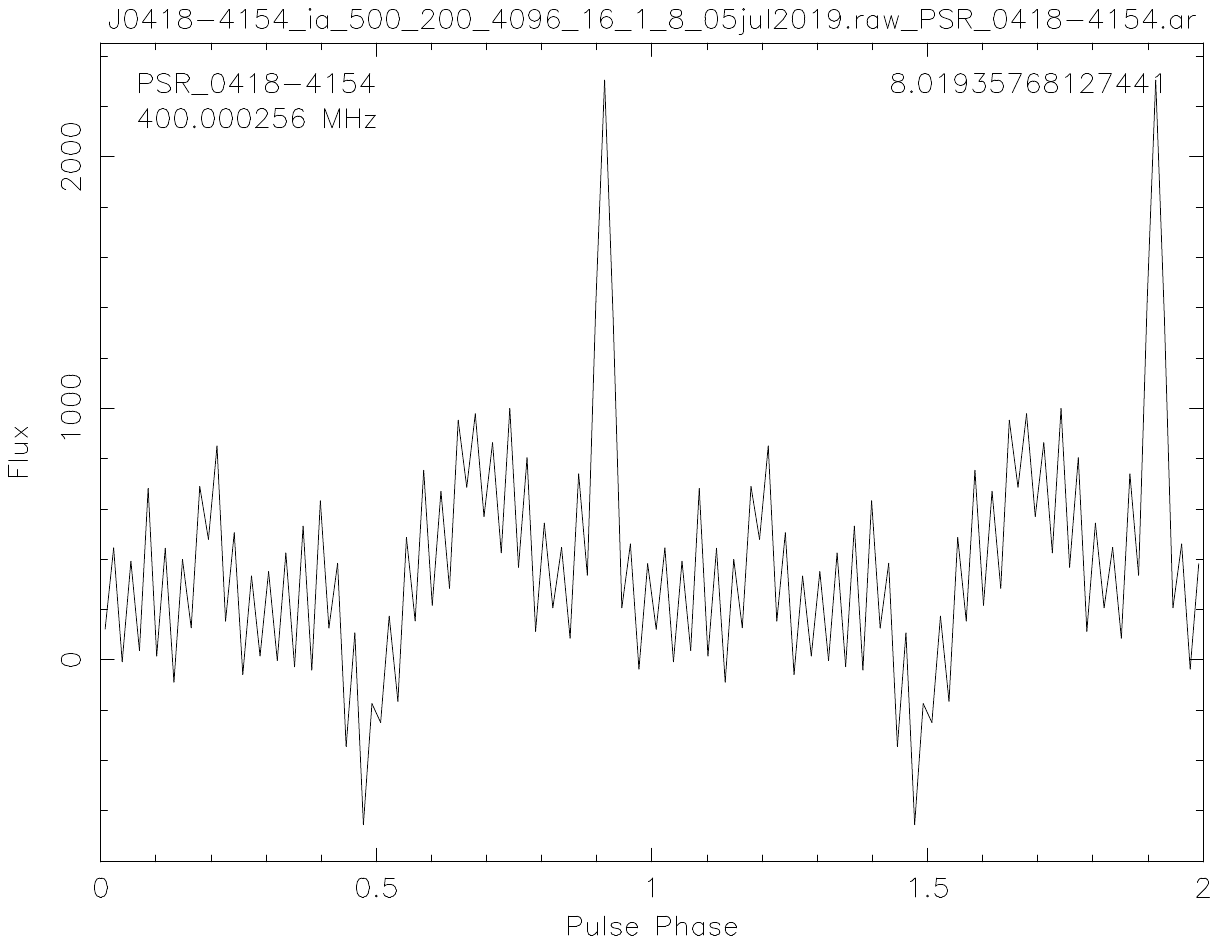}
 \centering

 \caption{Unfiltered profile of PSR J0418-4154.}
 \label{psrju}
 \endminipage
\hspace{0.05\linewidth}
\minipage{0.45\linewidth}
\includegraphics[scale=0.345]{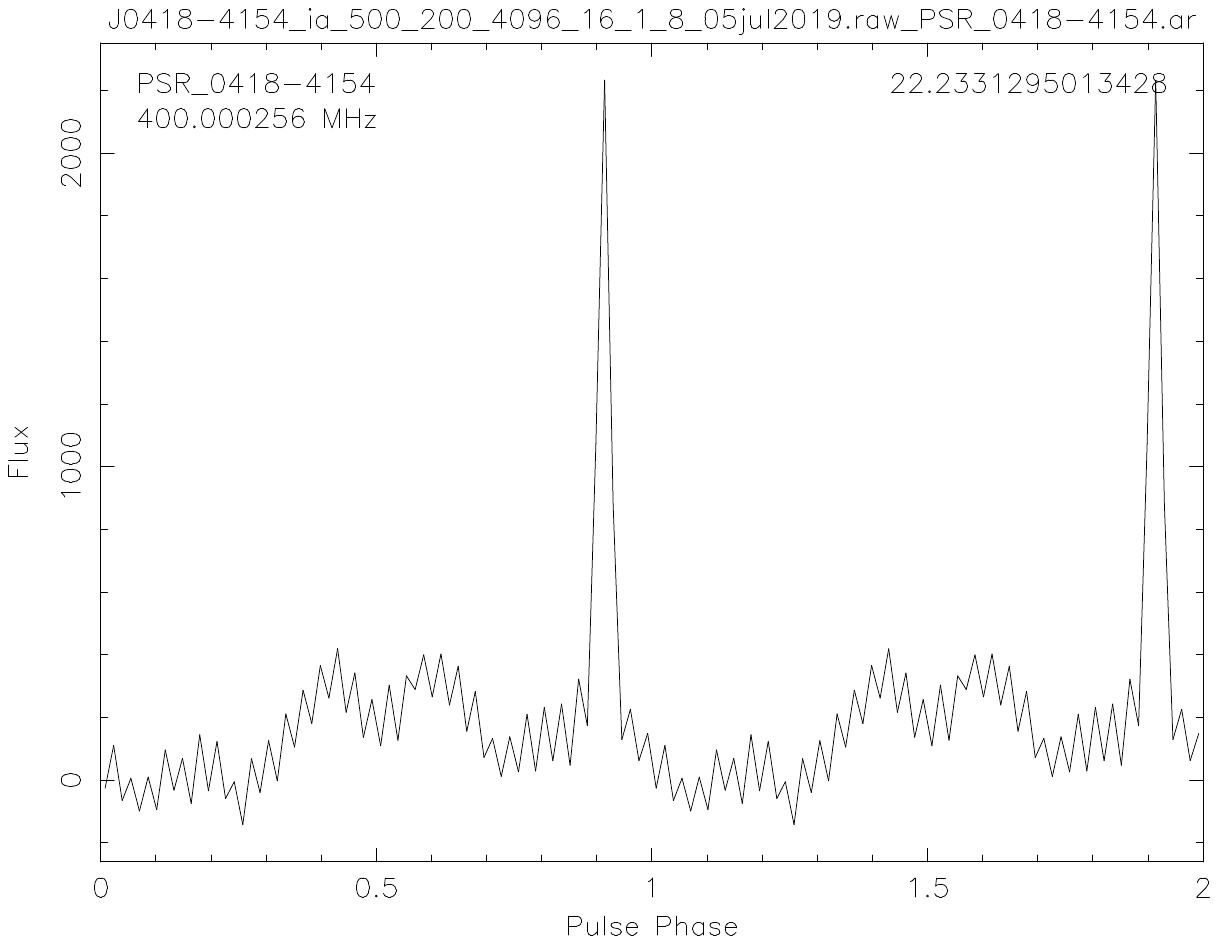}

 \centering
 \caption{Filtered profile of PSR J0418-4154. Three orders of improvement in the SNR of the folded profile.}
 \label{psrjf}
 \endminipage
\end{figure}

 Similar observation on PSR J0418-4154, a 
much weaker one than B0329+54 (by three order of magnitude) also shows similar improvement 
in the mean SNR of the folded profile (Fig. \ref{psrju} and Fig. \ref{psrjf}). 

\begin{figure}[h]
\minipage{0.45\linewidth}
 \includegraphics[scale =0.345]{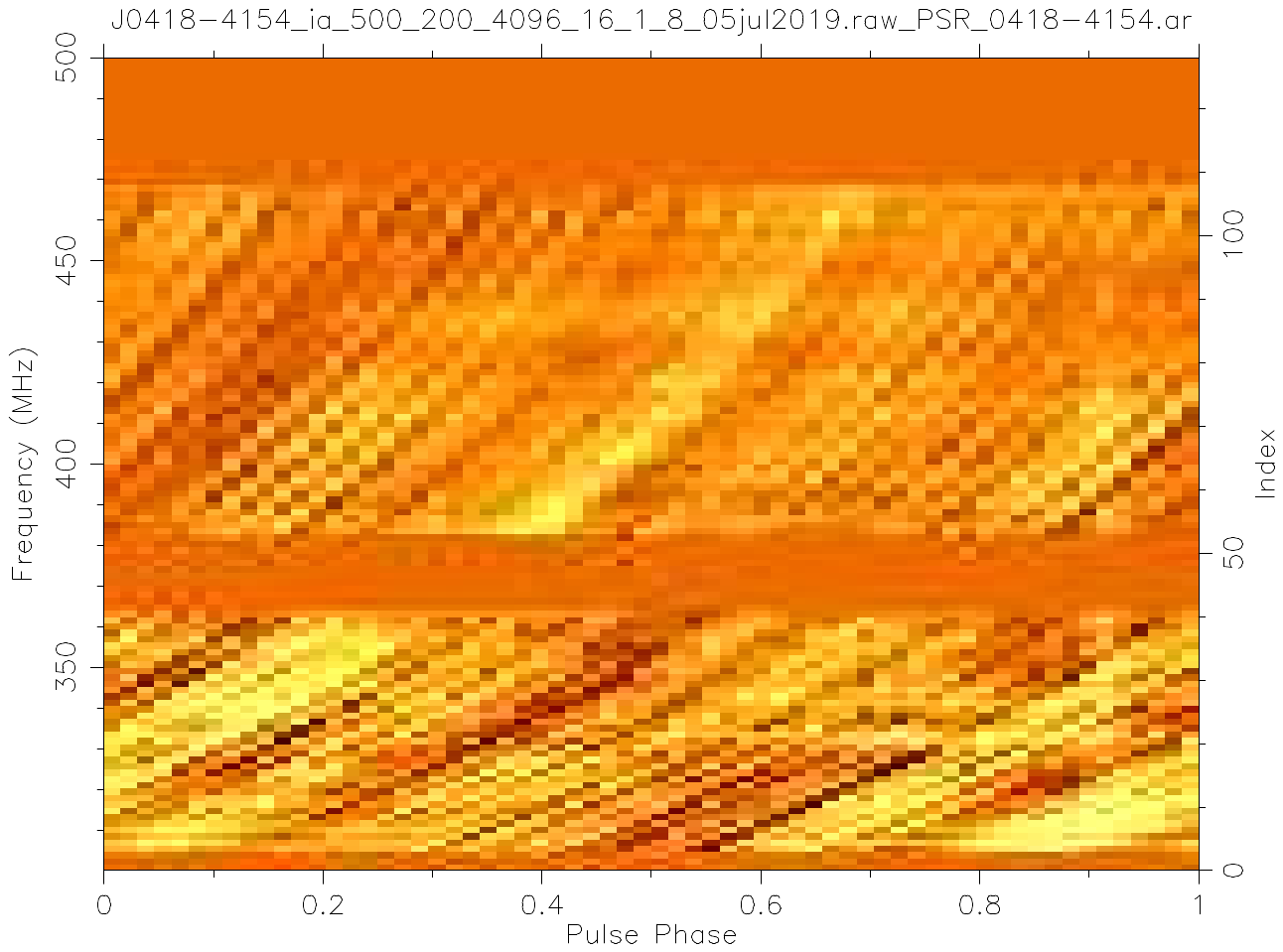}
 \centering

 \caption{Unfiltered frequency-phasogram for PSR J0418-4154.}
 \label{psrfpu}
 \endminipage
\hspace{0.05\linewidth}
\minipage{0.45\linewidth}
\includegraphics[scale=0.345]{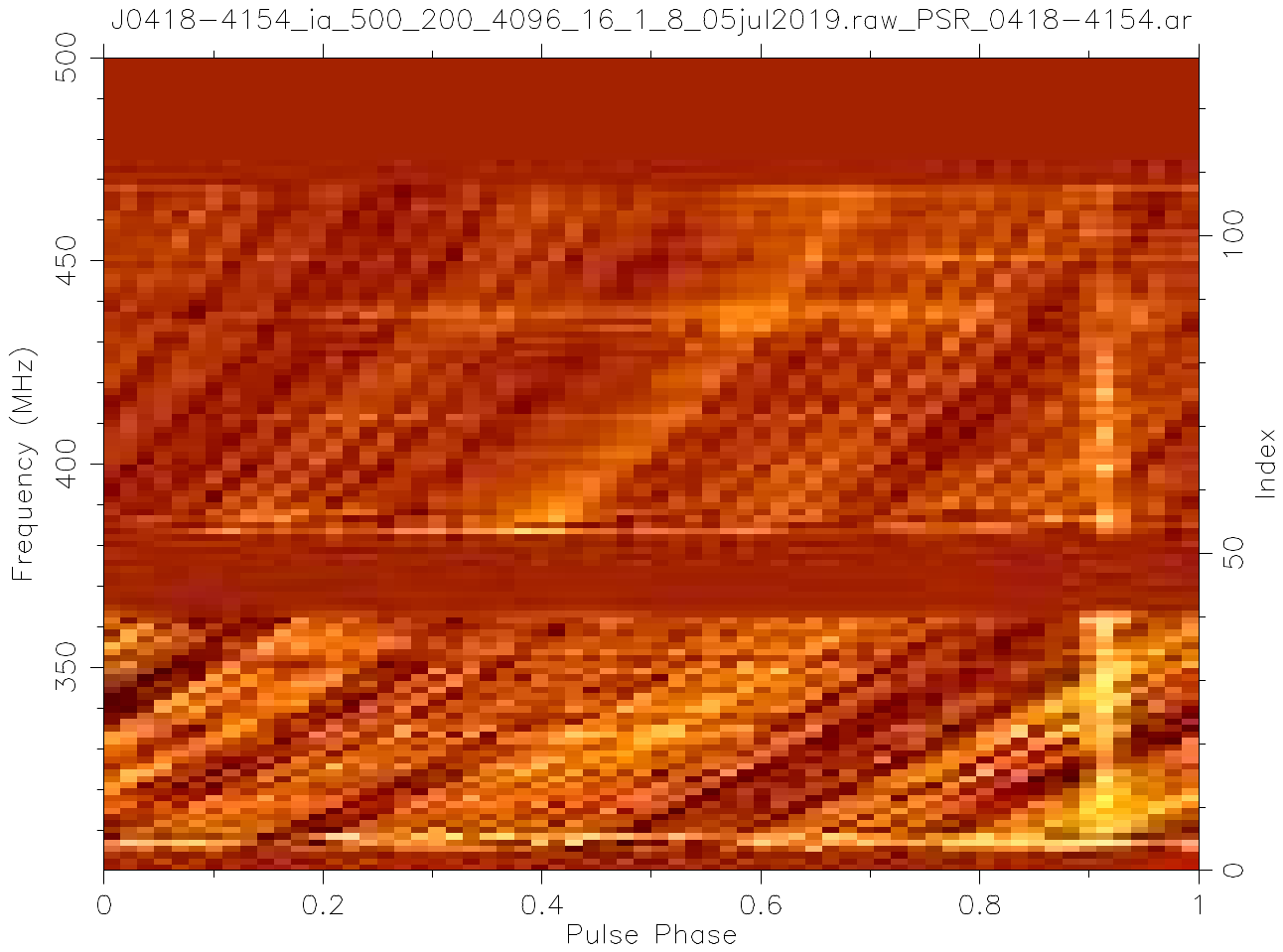}

 \centering
 \caption{Filtered frequency-phasogram for PSR J0418-4154.}
 \label{psrfpf}
 \endminipage
\end{figure}

The frequency-phasogram is also much cleaner for the
filtered data (Fig. \ref{psrfpu} and Fig. \ref{psrfpf}).

\clearpage

 \newpage
 \section{Results from interferometric mode tests}

A summary of the tests in the interferometric mode is provided in Table~\ref{testobs}.
\begin{table}[h]
\centering
\caption{Summary of the imaging tests using the voltage detection method for online RFI excision.}\label{testobs}
\vspace*{0.2cm}
 \begin{tabular}{ccccc}
 \hline
 \hline
  Date           & Band (Freq.) & Mode              & Sources & Time (min)\\
  \hline
  14 Jun. 2018  &  4 (610 MHz)  & 1:2               & 3C48 & 20\\
  11 Sept. 2018  & 2 (150 MHz)   & 1:2               & 0319+415(3C84),0432+416, 3C48  &80, 20, 10\\
  11 Aug. 2019   & 3 (330 MHz)   & 1:2               & A2744, 0116-208, 3C48&180, 35, 20\\
  13 Aug. 2019   & 4 (610 MHz)   & 1:2               & A2744,0116-208, 3C48& 240, 45, 30\\
  09 Sept. 2019  & 5 (1400 MHz) & 1:2               & 3C147 & 35\\
  \hline
   \hline
 \end{tabular}
\end{table}



\subsection{Effect of filtering on the visibilities}\label{viseff}
In this section, results from the test in 1:2 mode at band 4 (14 June 2018) are described as an example of the differences expected in data with and without filtering. The data for this test were analysed in AIPS. The filtered and unfiltered data were processed identically. 
\subsubsection{UV-distance and frequency}
\begin{figure}[h]
\begin{center}
        \includegraphics[trim={1cm 4cm 1cm 3cm}, clip,height=8cm]{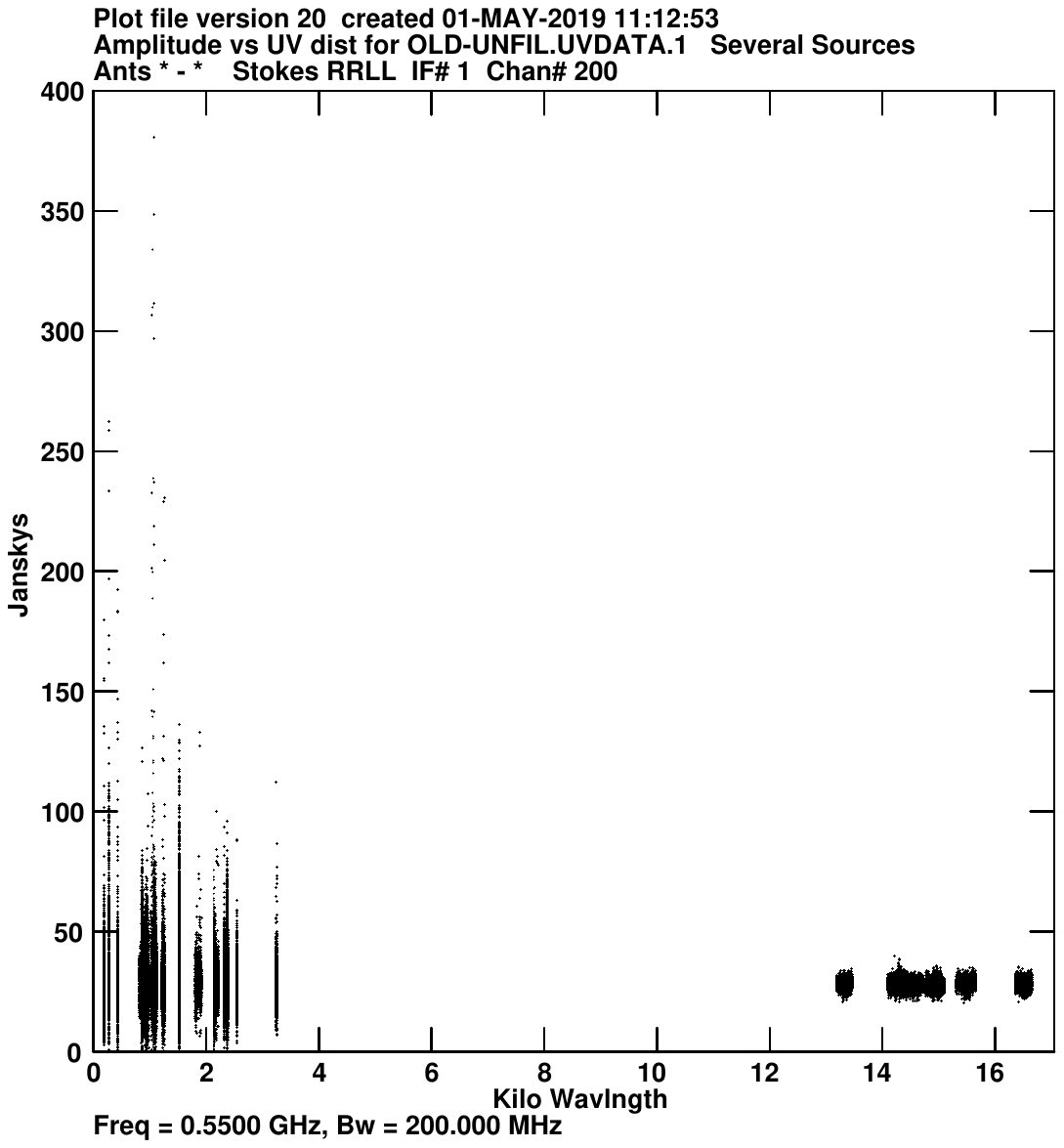}
        \includegraphics[trim={1cm 4cm 1cm 3cm}, clip,height=8cm]{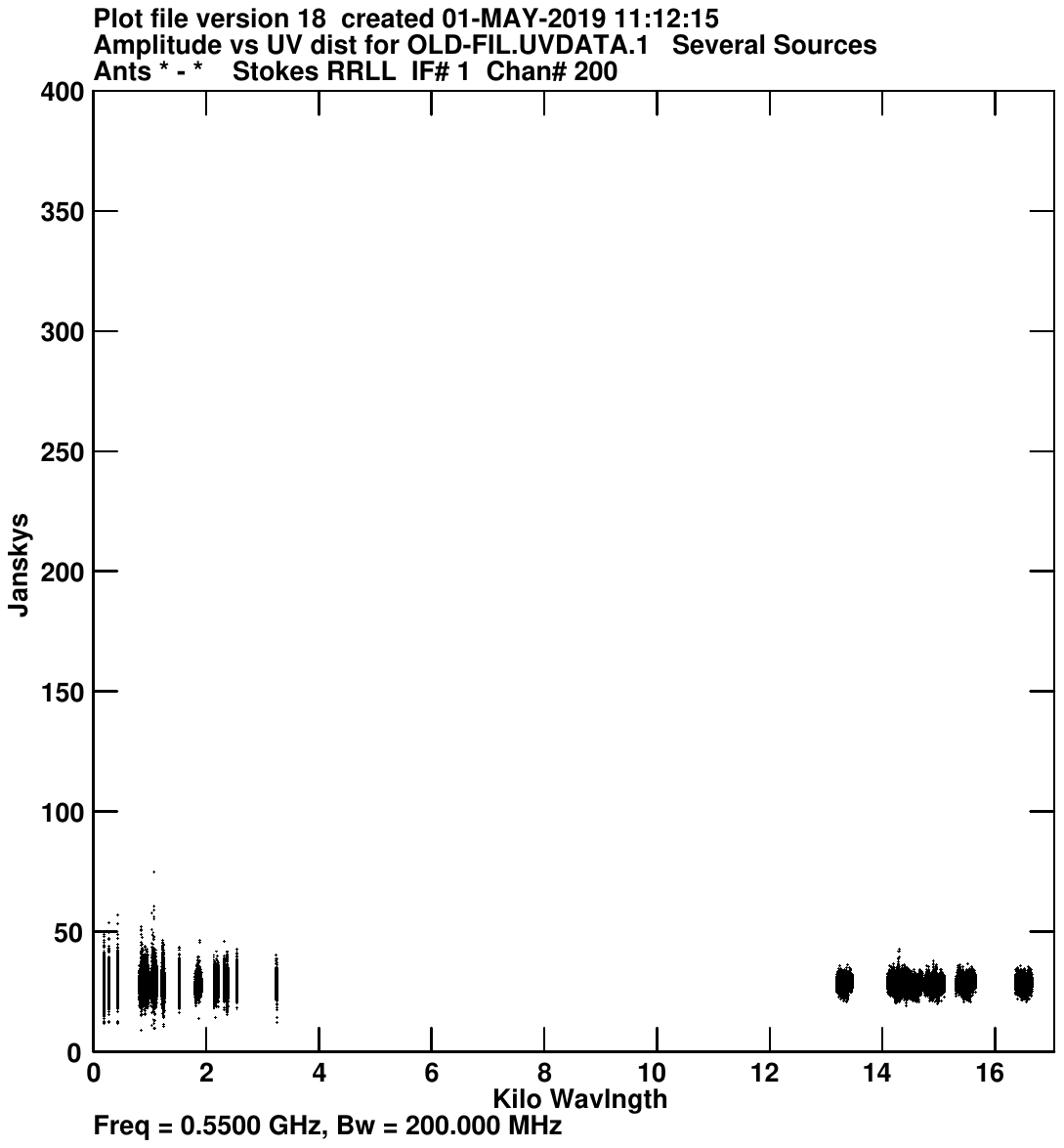}
 \caption{Band 4, 14 June 2018: Calibrated amplitude versus uv-distance is shown for the unfiltered (left) and filtered (right) data 
 for channel 200 (0.550 GHz). The short baselines where the BB RFI is known to be correlated show significant improvement. [Sec.~\ref{viseff}]
}
 \label{14june-uvplt}
 \end{center}
\end{figure}

\begin{figure}[h]
\begin{center}
        \includegraphics[trim={0cm 8cm 0cm 0cm}, clip,height=7.0cm]{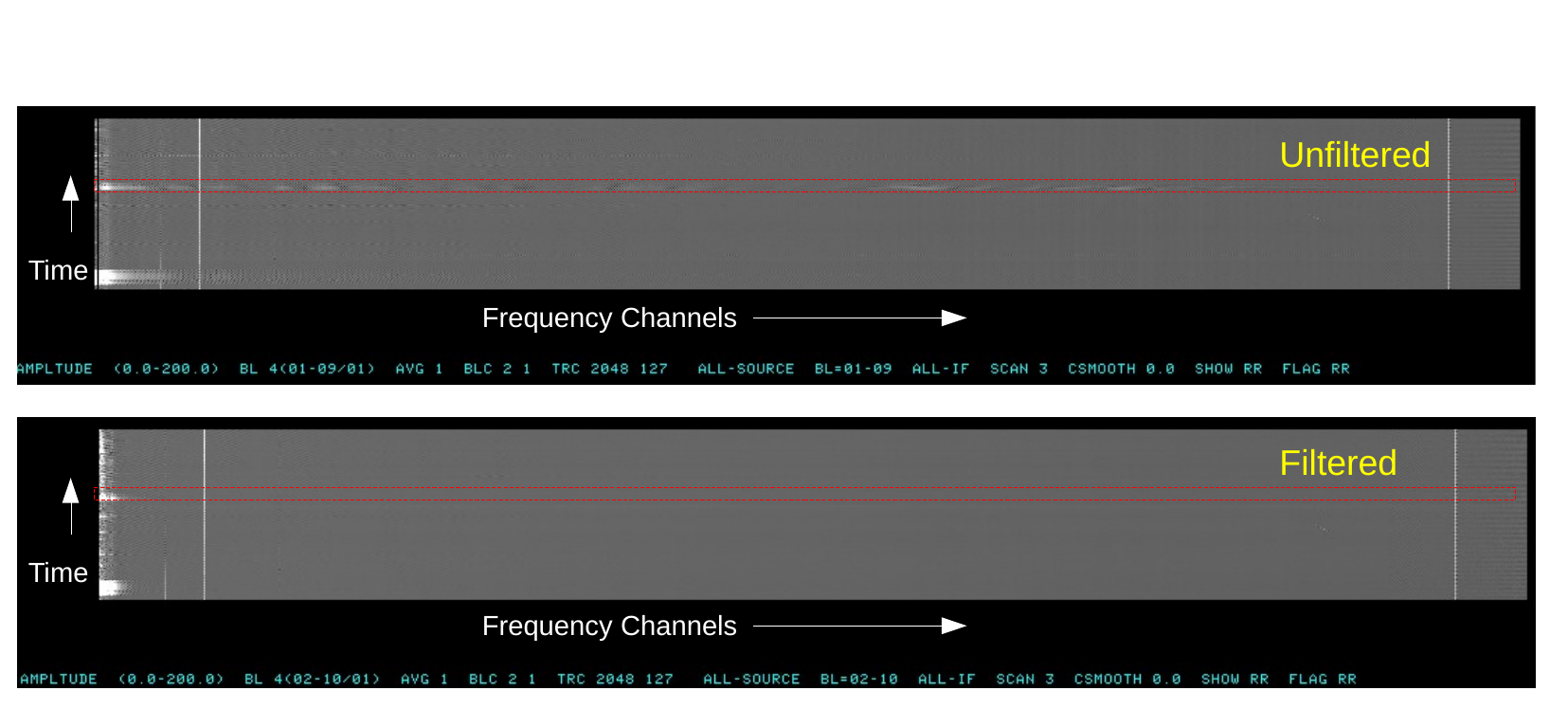}
 \caption{Band 4, 14 June 2018: Time-frequency plane for the baseline C00-C09 in unfiltered (C01-C10 in filtered) is shown on the same grey-scale ranging from 0 - 200 Jy. A broadband feature is marked with a box in both unfiltered and filtered data. [Sec.~\ref{viseff}]
}
 \label{14june-spflg}
 \end{center}
\end{figure}

\begin{figure}[h]
\begin{center}
        \includegraphics[trim={1cm 15cm 1cm 0cm}, clip,height=6cm]{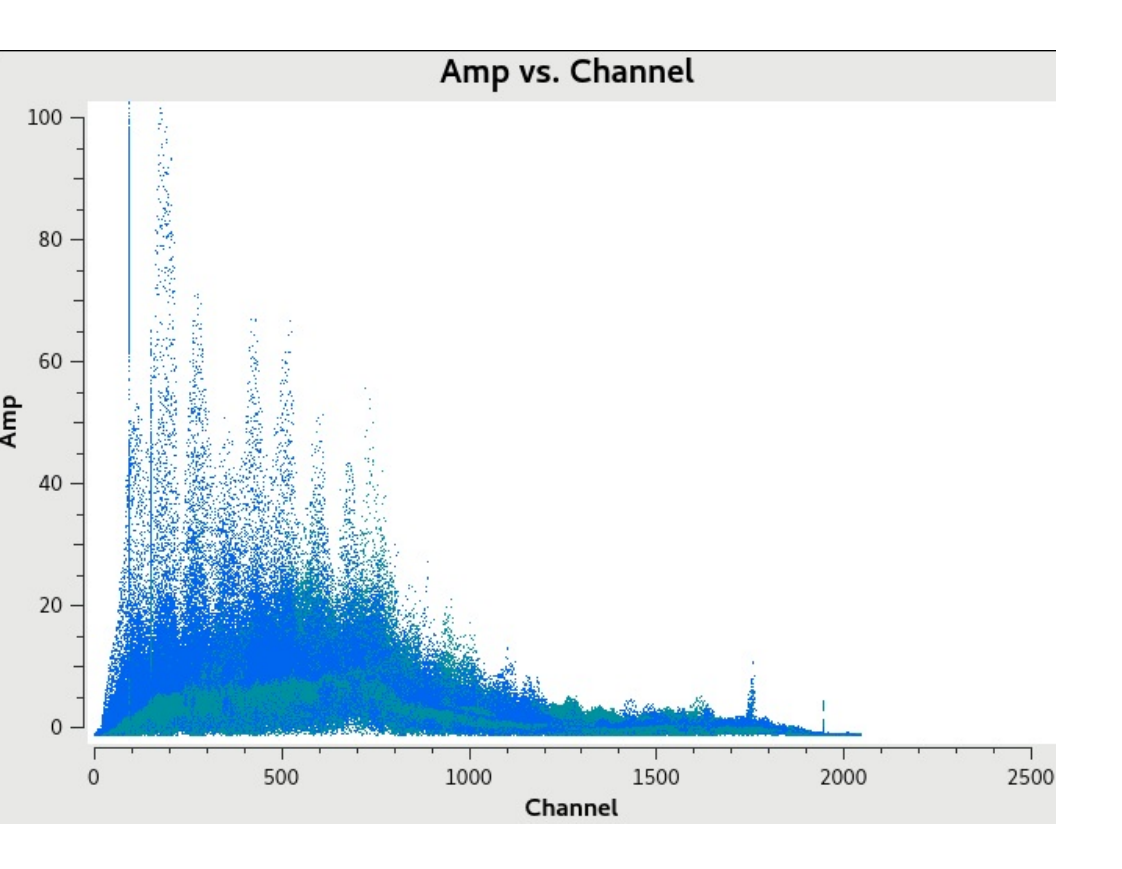}
        \includegraphics[trim={1cm 15cm 1cm 0cm}, clip,height=6cm]{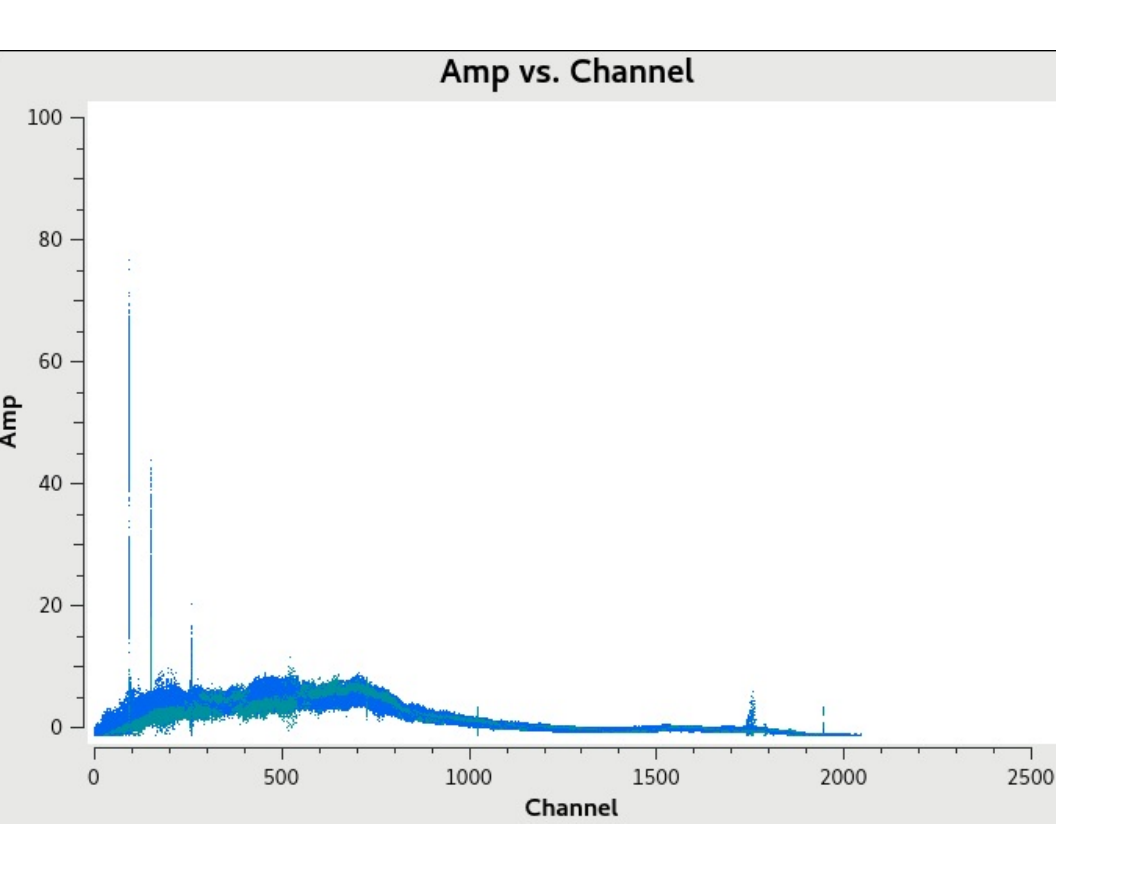}
 \caption{Band 4, 14 June 2018: The uncalibrated spectrum of the baseline C00-C02 in unfiltered data is shown in the left column (filtered copy C01-C03 in the right column). The LL polarization is shown in blue and RR in green. [Sec.~\ref{viseff}]
}
 \label{14june-spec}
 \end{center}
\end{figure}

In Fig.~\ref{14june-uvplt}, the calibrated amplitude is plotted as a function of uv-distance for channel 200. The difference due to the filter is mainly evident at the short baselines where the number of outlier points in unfiltered data are seen. 
 The time-frequency plane with amplitude plotted as a grey-scale is shown in Fig.~\ref{14june-spflg}. NB and BB RFI are both seen, and the effect on one 
broadband feature is highlighted. An example of an uncalibrated spectrum is shown in Fig.~\ref{14june-spec}. The two polarizations are shown in two colours. The large scatter on the amplitudes across the spectrum in the unfiltered data are not present in the filtered data. Such effects of broadband RFI cannot be removed by flagging in the post-processing of data and must be done with a real time filtering.

\begin{figure}[h]
\begin{center}
        \includegraphics[trim={1cm 3cm 1cm 3cm}, clip,height=6.5cm]{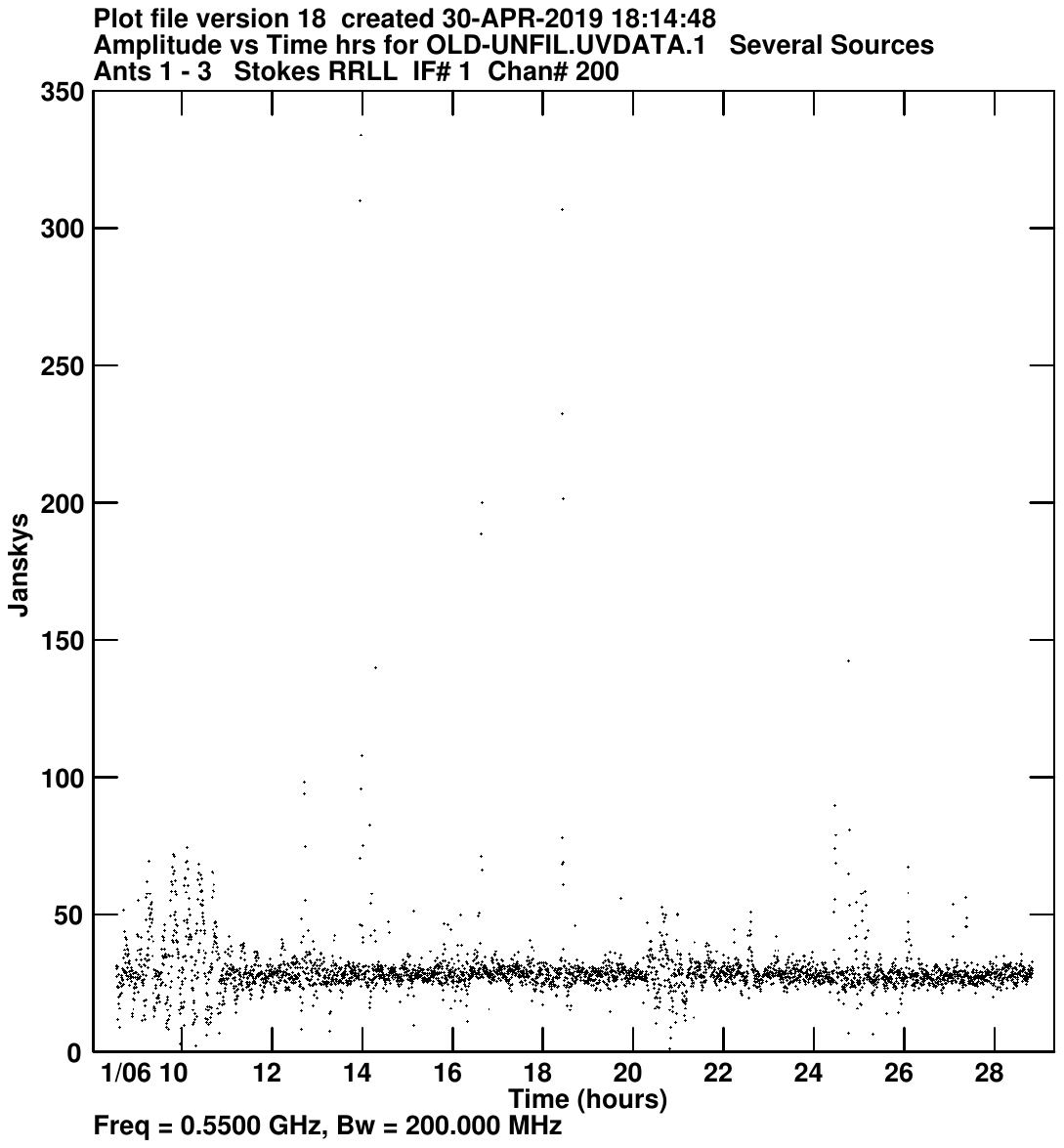}
 \includegraphics[trim={1cm 3cm 1cm 3cm}, clip,height=6.5cm]{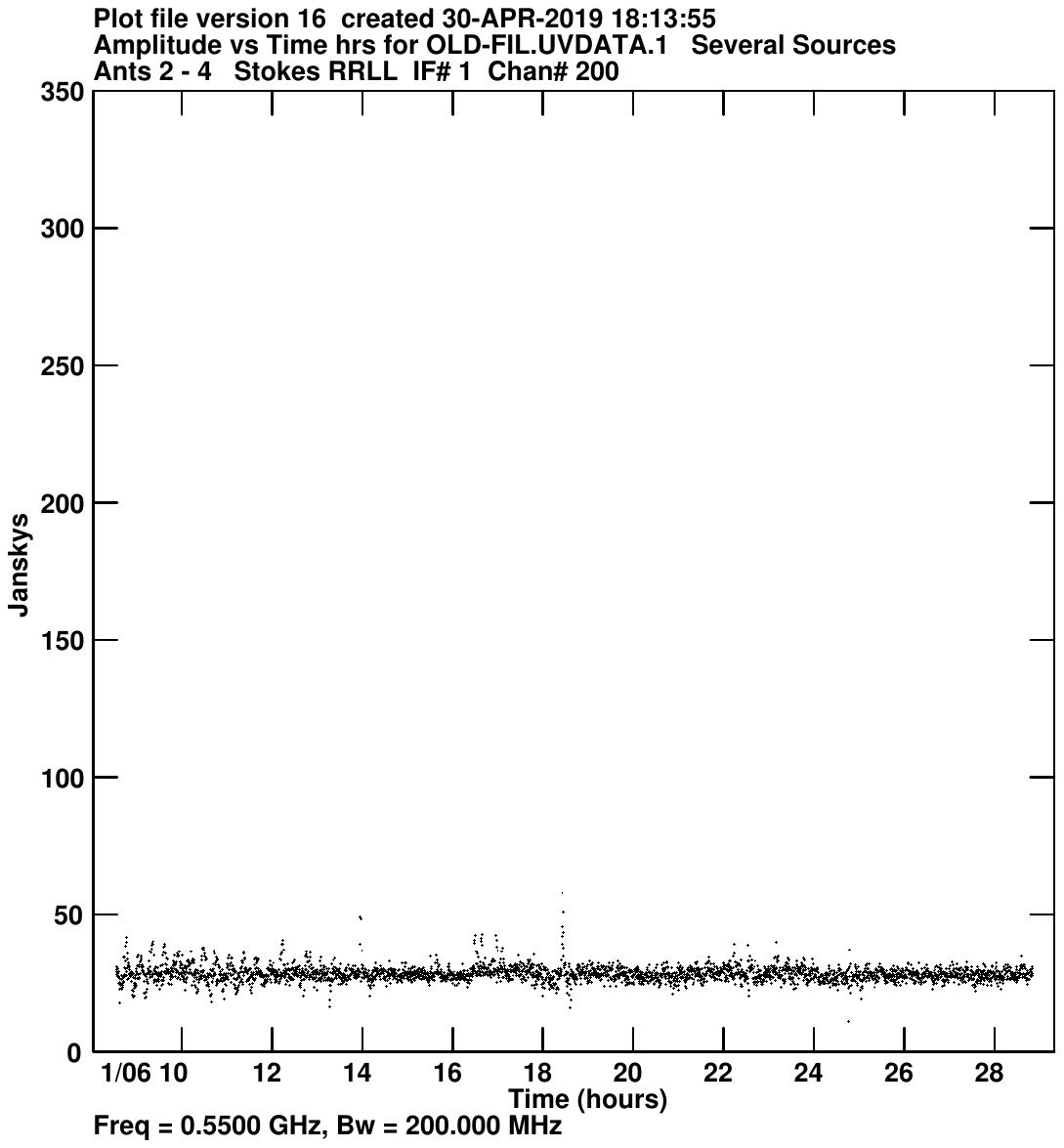}\\
 \includegraphics[trim={1cm 3cm 1cm 3cm}, clip,height=6.5cm]{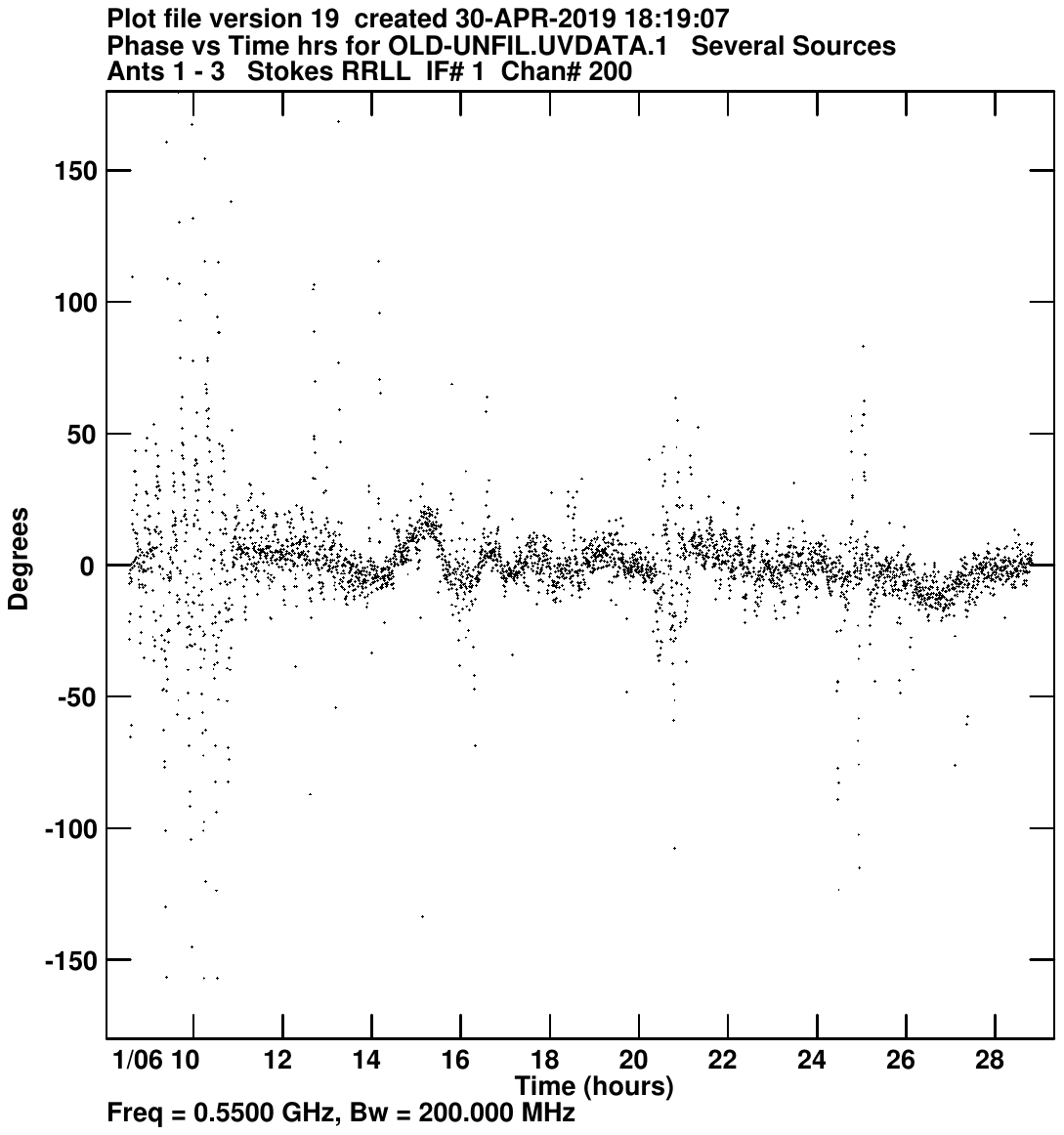}
  \includegraphics[trim={1cm 3cm 1cm 3cm}, clip,height=6.5cm]{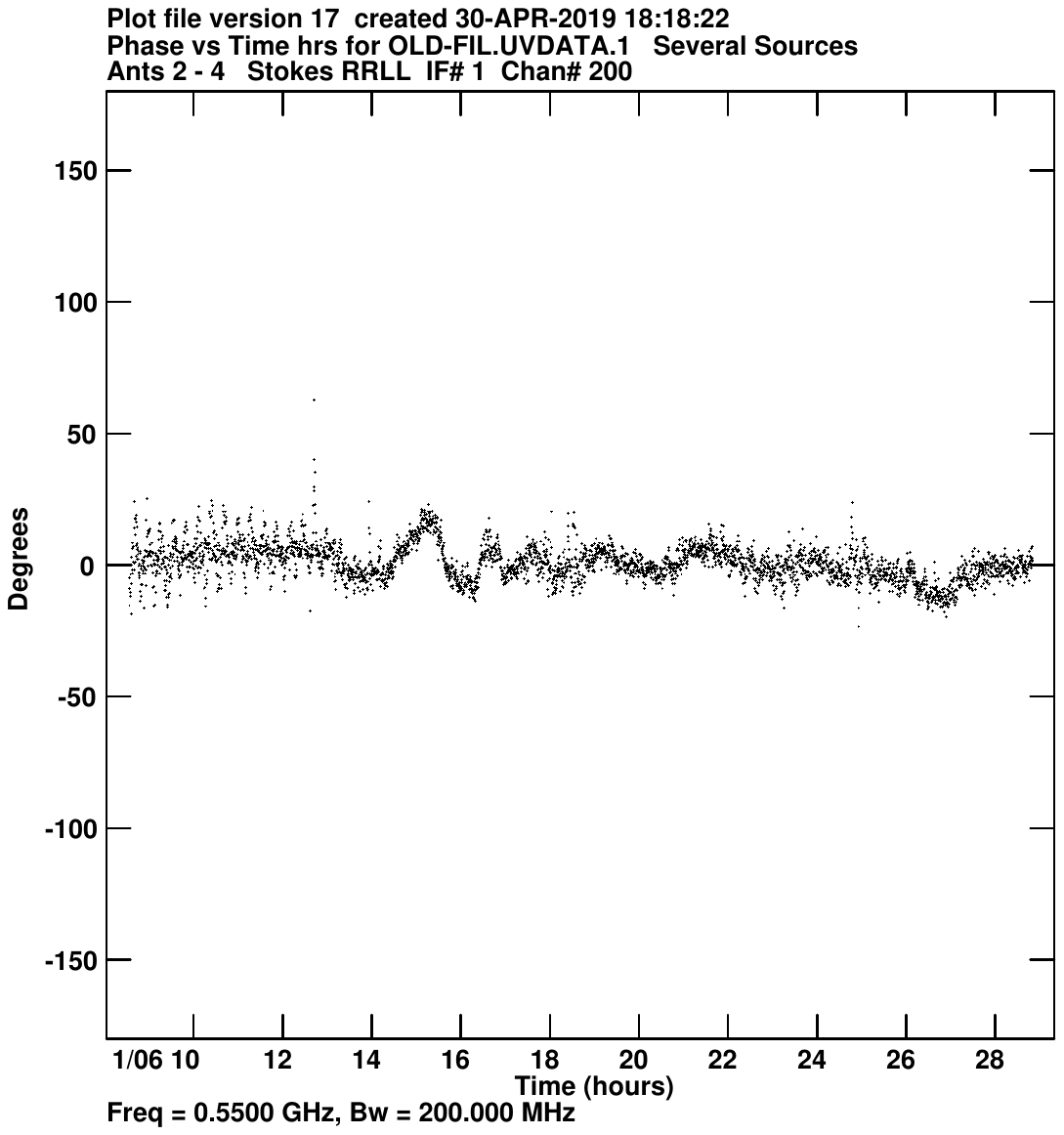}\\
    \includegraphics[trim={1cm 3cm 1cm 3cm}, clip,height=6.5cm]{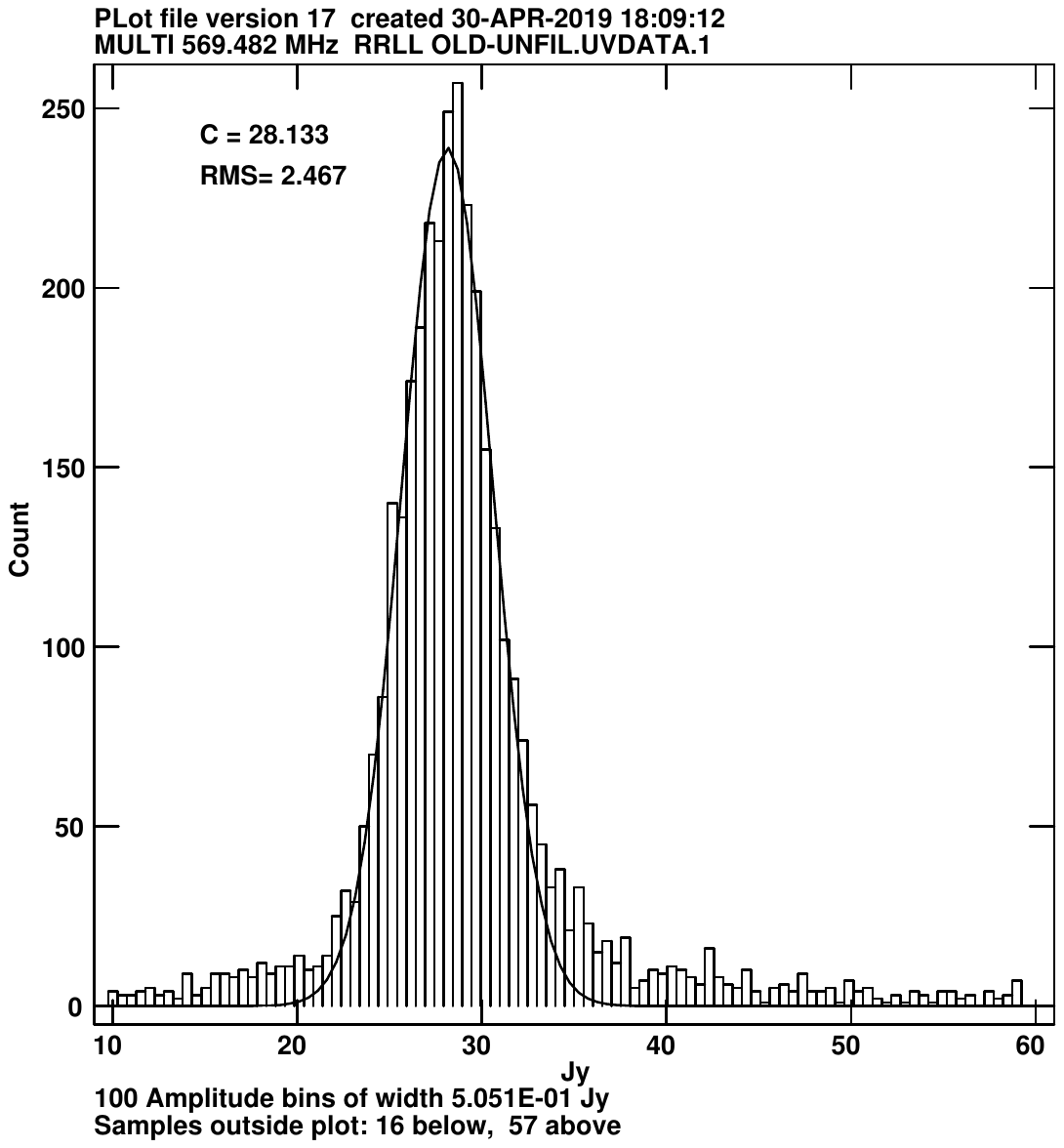}
    \includegraphics[trim={1cm 3cm 1cm 3cm}, clip,height=6.5cm]{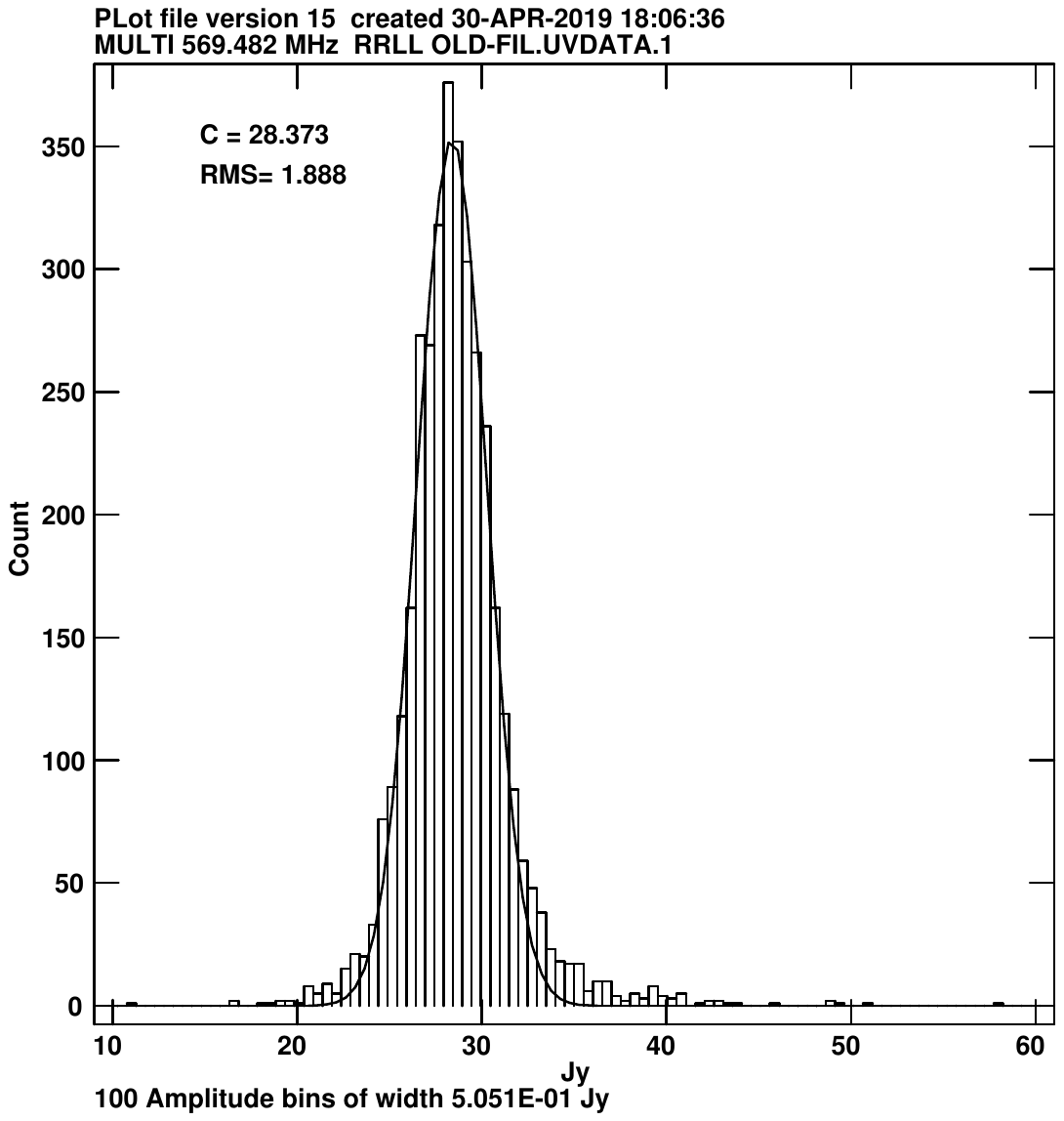}
 \caption{Band 4, 14 June 2018: The baseline C00-C02 in unfiltered data is shown in the left column (filtered copy C01-C03 in the right column) for channel 200.
 The top panels show Amplitude Vs time, and the middle panels show Phase vs. time. The bottom panels show the histograms of the amplitudes with a Gaussian fit 
 for unfiltered and filtered data. [Sec.~\ref{viseff}]}
 \label{14june-time}
 \end{center}
\end{figure}

The amplitude and phase as functions of time for a single baseline after calibration for unfiltered and filtered are shown in the top and middle panels of Fig.~\ref{14june-time}. The effect of filtering is to reduce the scatter along the y-axis in the two plots. 
 Histograms of the calibrated visibility amplitudes in unfiltered and filtered cases are shown in the bottom panels of Fig.~\ref{14june-time}. The thickness of the tails of the histogram is reduced in the filtered data as can be seen in the reported rms on the peak of 2.5 Jy in unfiltered versus 1.9 Jy in the filtered data. 

\subsubsection{Visibility amplitude standard deviation}
The online RFI excision replaces samples that cross the threshold with digital noise. BB RFI is more correlated with central square baselines as 
compared to the arm antennas. We expect that the RFI filtering mostly removes the RFI, and the effect of replacement by digital noise is very small. However, it may start to 
affect the actual correlation from the sky. We examined the standard deviations on the calibrated amplitudes of 
calibrator data as a function of baseline length and quantified the effect when a threshold of $3\sigma$ and replacement with digital noise was used for the filtering. 

\begin{figure}[h]
\begin{center}
        \includegraphics[trim={3cm 1.5cm 1.5cm 2cm}, clip,height=7.0cm,angle=90]{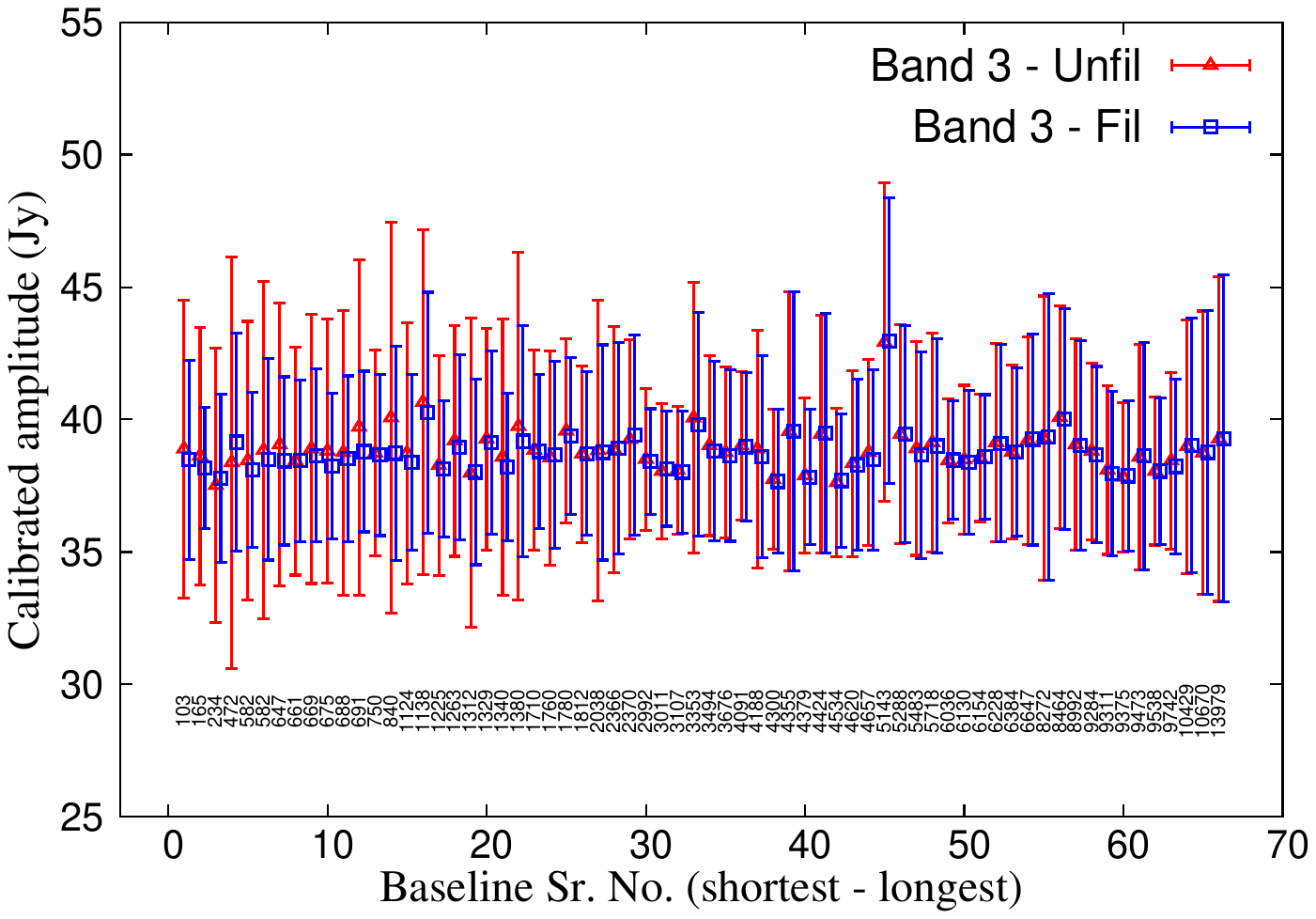}
         \includegraphics[trim={3cm 1.5cm 1.5cm 2cm}, clip,height=7.0cm,angle=90]{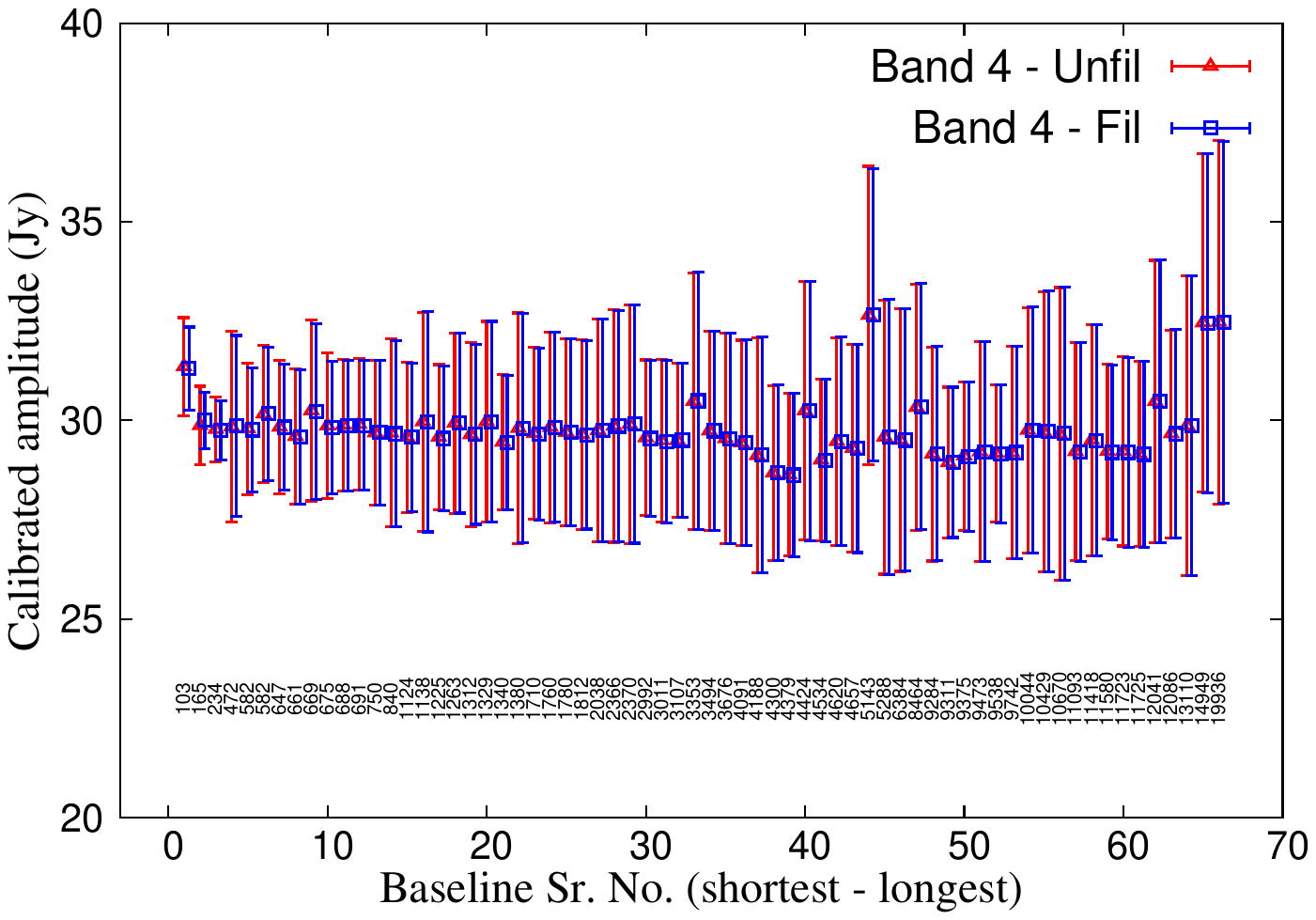}
         \includegraphics[trim={3cm 1.5cm 1.5cm 2cm}, clip,height=7.0cm,angle=90]{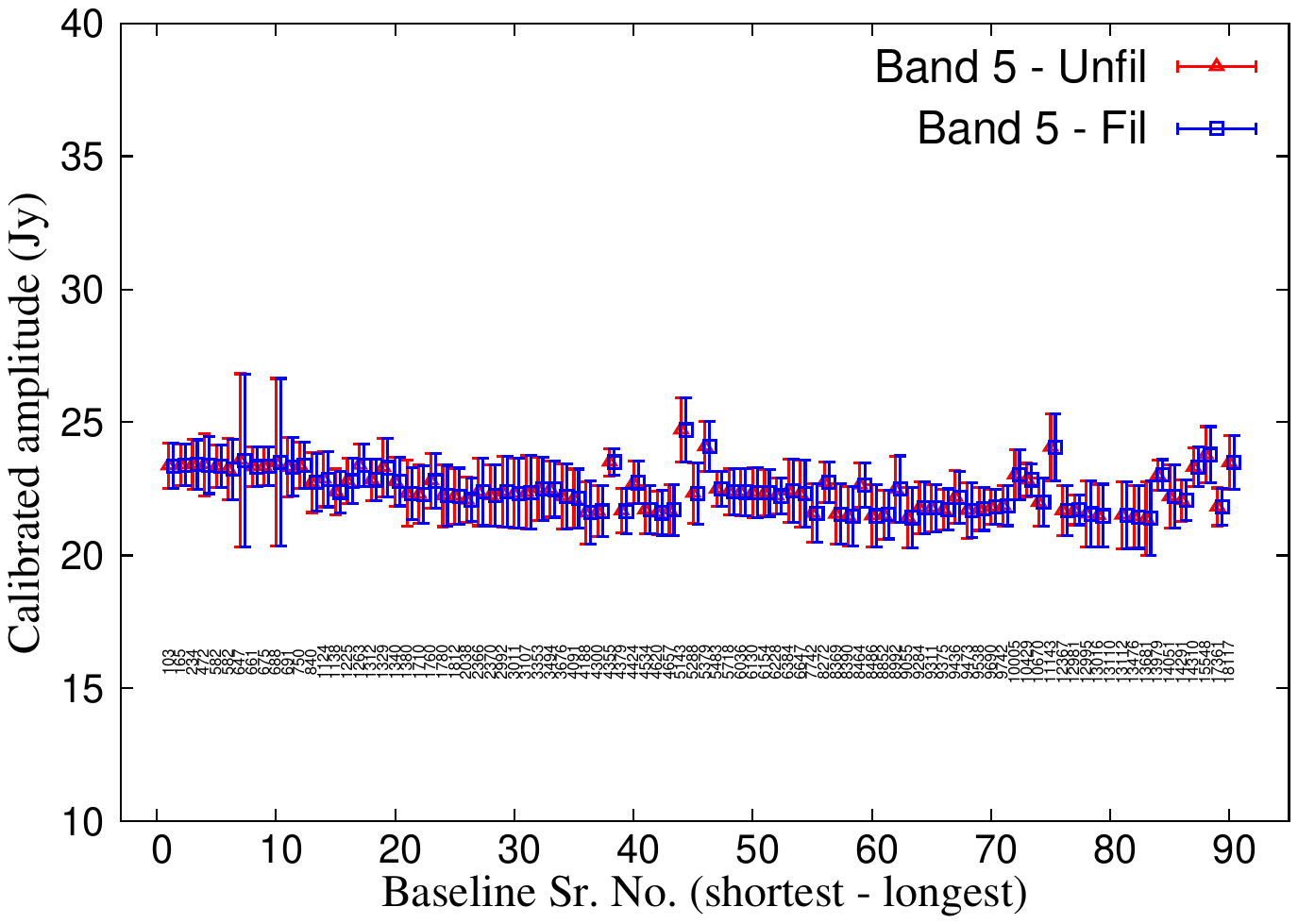}
 \caption{The calibrated flux density of a primary calibrator source for a single channel is plotted as a function of baseline numbers sorted according to their lengths. The integers in the plot 
 indicate the length of the baseline in metres, ranging from $100$ m to 14, 20 and 15.5 km for bands 3, 4 and 5, respectively. The unfiltered (red) and filtered (blue) points are offset on the x-axis for comparison. [Sec.~\ref{viseff}]}
 \label{b3blplot}
 \end{center}
\end{figure}

In Fig.~\ref{b3blplot}, the unfiltered and filtered data are shown. For band 3, at short baselines, the effect of the filter to reduce the standard deviation of 
the amplitude is seen. The effect of filtering is less at bands 4 and 5, as these bands are less affected by BB RFI. 

\begin{figure}[h]
\begin{center}
        \includegraphics[trim={1cm 2.5cm 1cm 2cm}, clip,height=7.0cm]{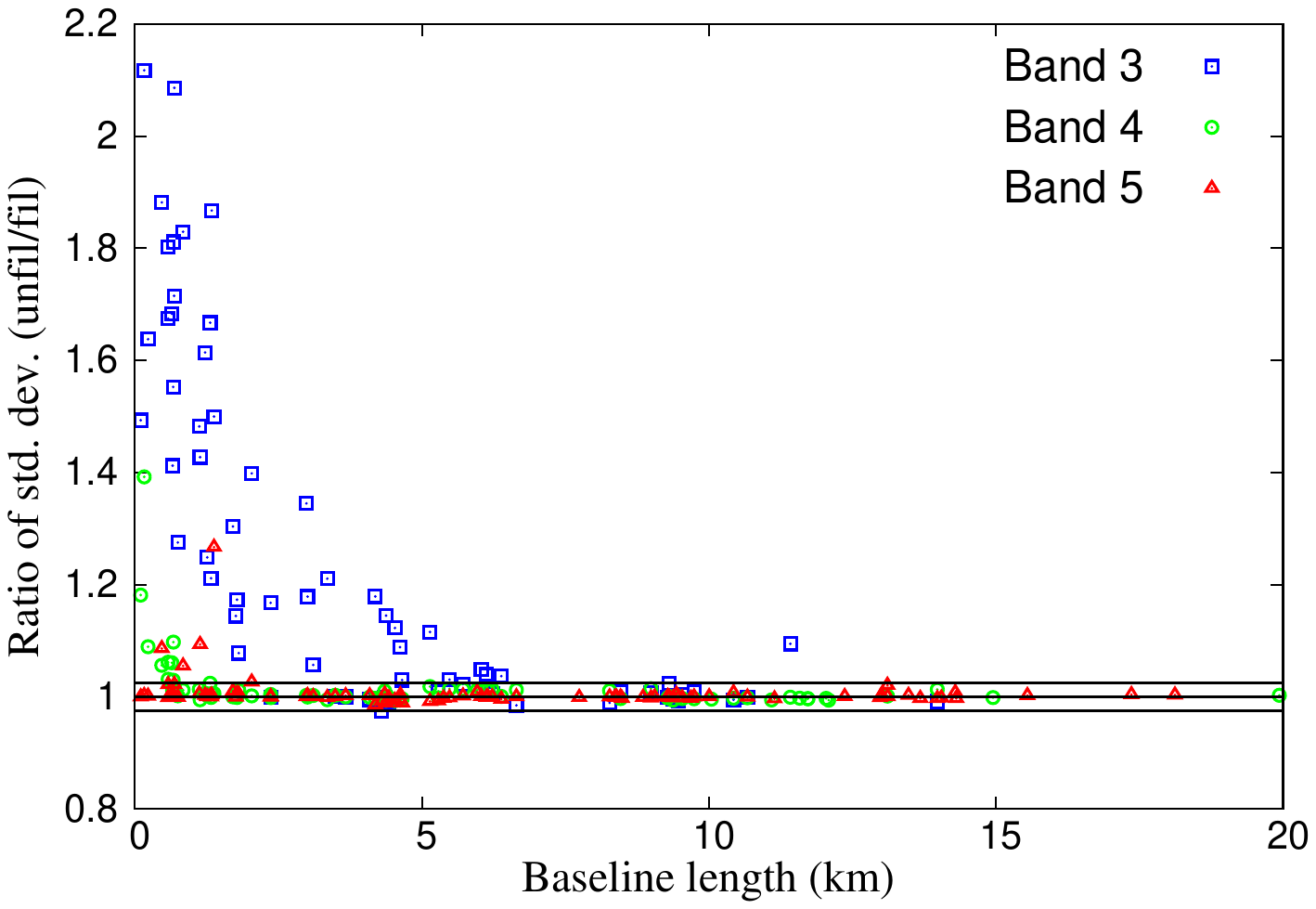}
 \caption{The ratio of standard deviations on the visibility amplitude (unfiltered/filtered) versus baseline length is shown. At short baselines, there is considerable improvement 
 , and at long baselines it is close to 1. This quantity is plotted for bands 3, 4 and 5. For a few baselines, the filtered standard 
 deviation is larger than that of unfiltered - this is confined to less than $2.5\%$. The horizontal lines are at 1.0, 1.025 and 0.975. [Sec.~\ref{viseff}]}
 \label{b345plot}
 \end{center}
\end{figure}

We compare the filtering 
for different bands in Fig.~\ref{b345plot}. The ratio of standard deviations in unfiltered and filtered data is plotted in serial order from left to right 
in ascending order of baseline length (also given in the plot). A higher value of the ratio indicates that the standard deviation has improved by that factor. 
We also have a few points that are just below 1, which means that in those baselines, the signal after filtering showed a larger standard deviation than 
the unfiltered signal. We note that this is limited to less than $2.5\%$ at bands 3, 4 and 5.

\subsection{Calibrator flux densities}\label{calflux}
We compared the calibrator flux densities in order to check for any systematic effect produced by the filtering. Using the task 'visstat' in CASA, the flux densities of the calibrators were obtained over 
about 100 channels ($9.7$ MHz bandwidth). The calibrated flux densities are within a few per cent of those obtained from unfiltered data (Table ~\ref{calsrcflux}).

\subsection{Effect on the flagging percentage}\label{flgimg}
\begin{figure}[h]
\begin{center}
       \includegraphics[trim={1.5cm 1cm 2.5cm 2.0cm}, clip,height=8cm]{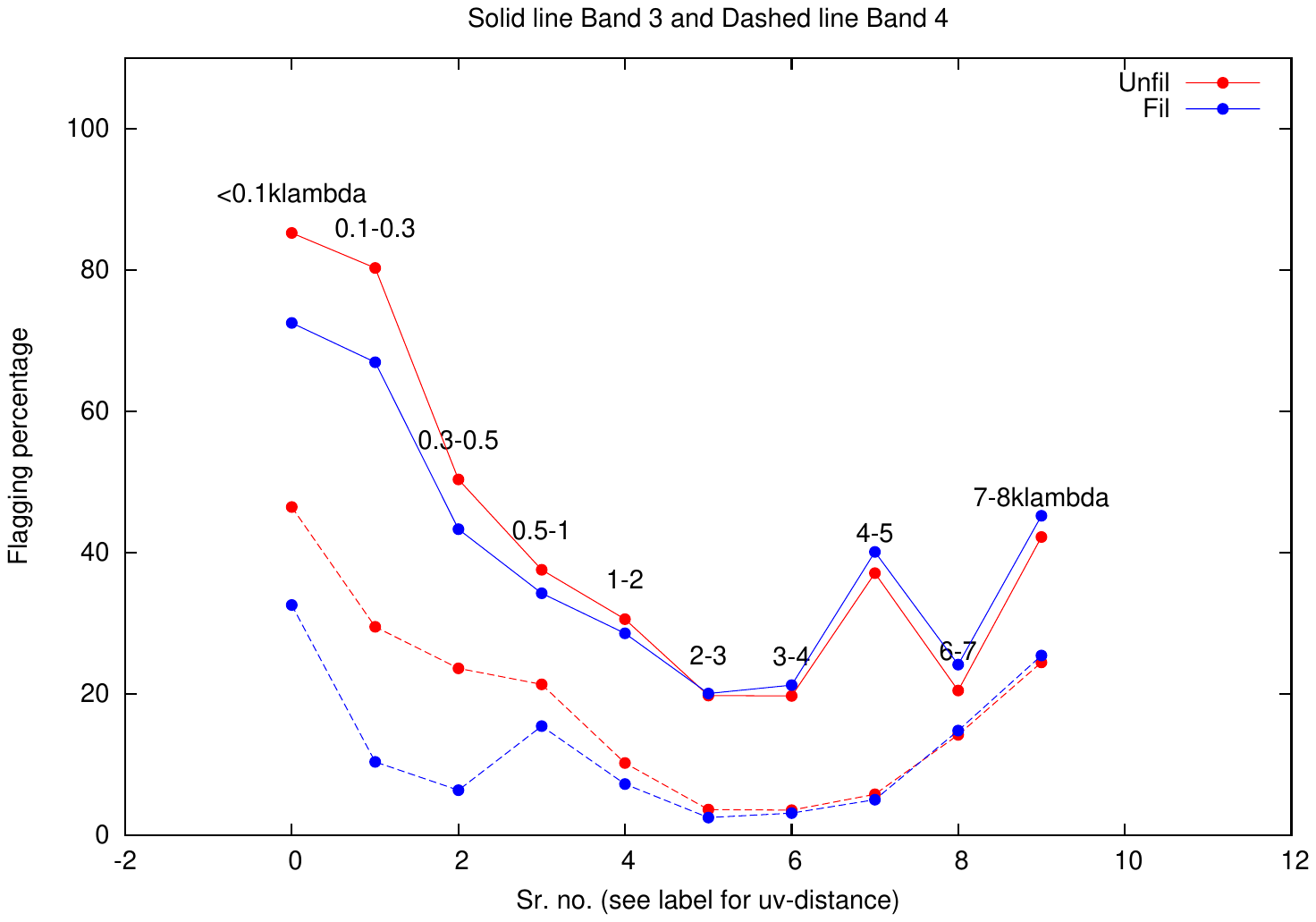}
 \caption{ Flagging percentage versus baseline serial numbers sorted according to their length are shown for bands 3 (solid) and 4 (dashed) using red colour for unfiltered data and blue for filtered data. The points are obtained in uv-distance bins as labelled above the points. [Sec.~\ref{flgimg}]}
 \label{flagpercent}
 \end{center}
\end{figure}

We processed the interferometric data using a pipeline in CASA. The pipeline was used to ensure that 
the filtered and unfiltered data were analysed identically. Standard steps of 
flagging, calibration and imaging were carried out. Self-calibration was carried out to improve the sensitivity. We compared the flagging statistics for the target source in the tests at bands 3 and 4 (Table~\ref{flagstat}).
At short baselines, the flagging is $4\%$ and $6\%$ higher in unfiltered data at bands 3 and 4, respectively. The change in the flagging percentage as a function of uv-distance is shown in Fig.~\ref{flagpercent}. At the short and intermediate length baselines, the percentage of data flagged is lower in the filtered case and comparable on longer baselines.

\subsection{Effect on the images}
\begin{figure}[h]
\begin{center}
        \includegraphics[trim={0cm 0cm 0cm 0cm}, clip,height=8.5cm]{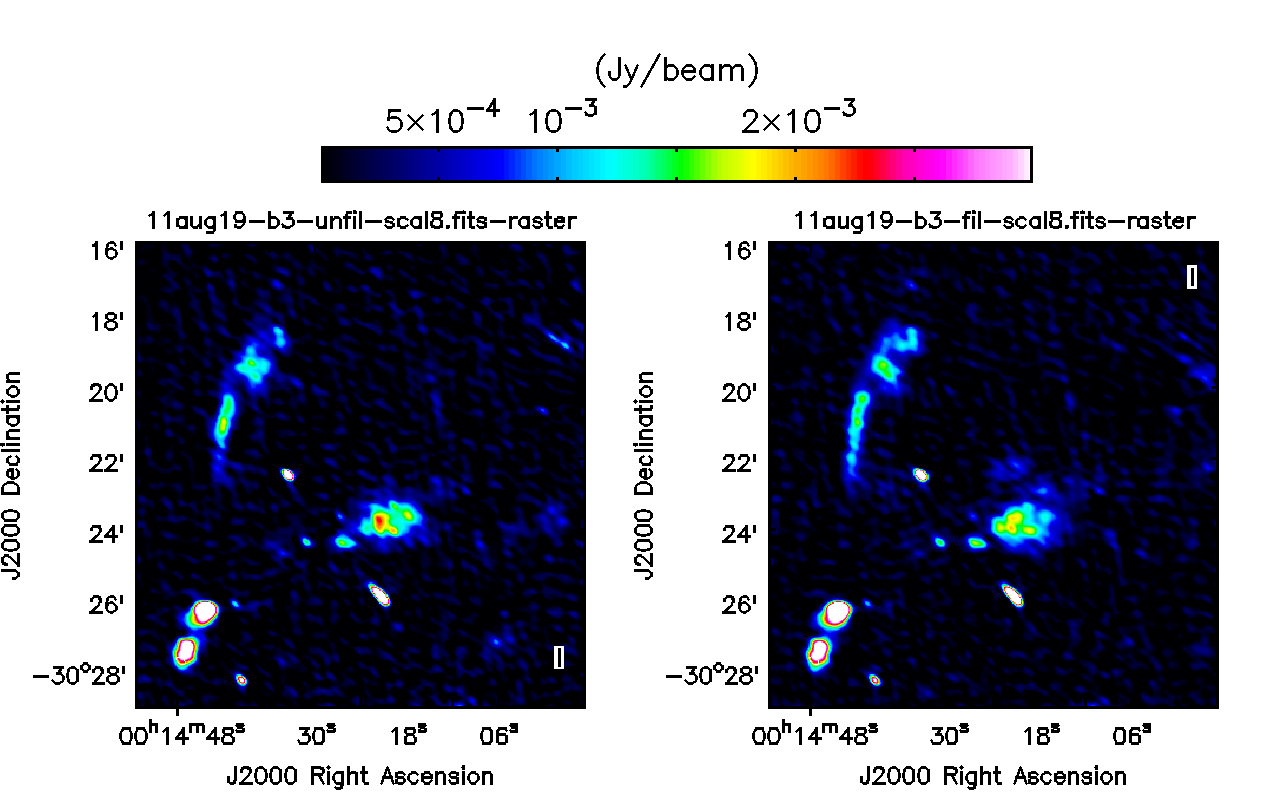}
 \caption{Band 3 images of A2744 field: unfiltered (left) and filtered (right). The rms in the unfiltered image is 115 $\mu$Jy beam$^{-1}$ and in the filtered image is 107 $\mu$Jy beam$^{-1}$. The beam size is $17.9''\times10.6''$, position angle $53^{\circ}$. The morphology of the extended emission has a smoother distribution in the filtered data as compared to the unfiltered data. [Sec.~\ref{flgimg}]
}
 \label{b3img}
 \end{center}
\end{figure}

\begin{figure}[h]
\begin{center}
        \includegraphics[trim={0cm 0cm 0cm 0cm}, clip,height=8.5cm]{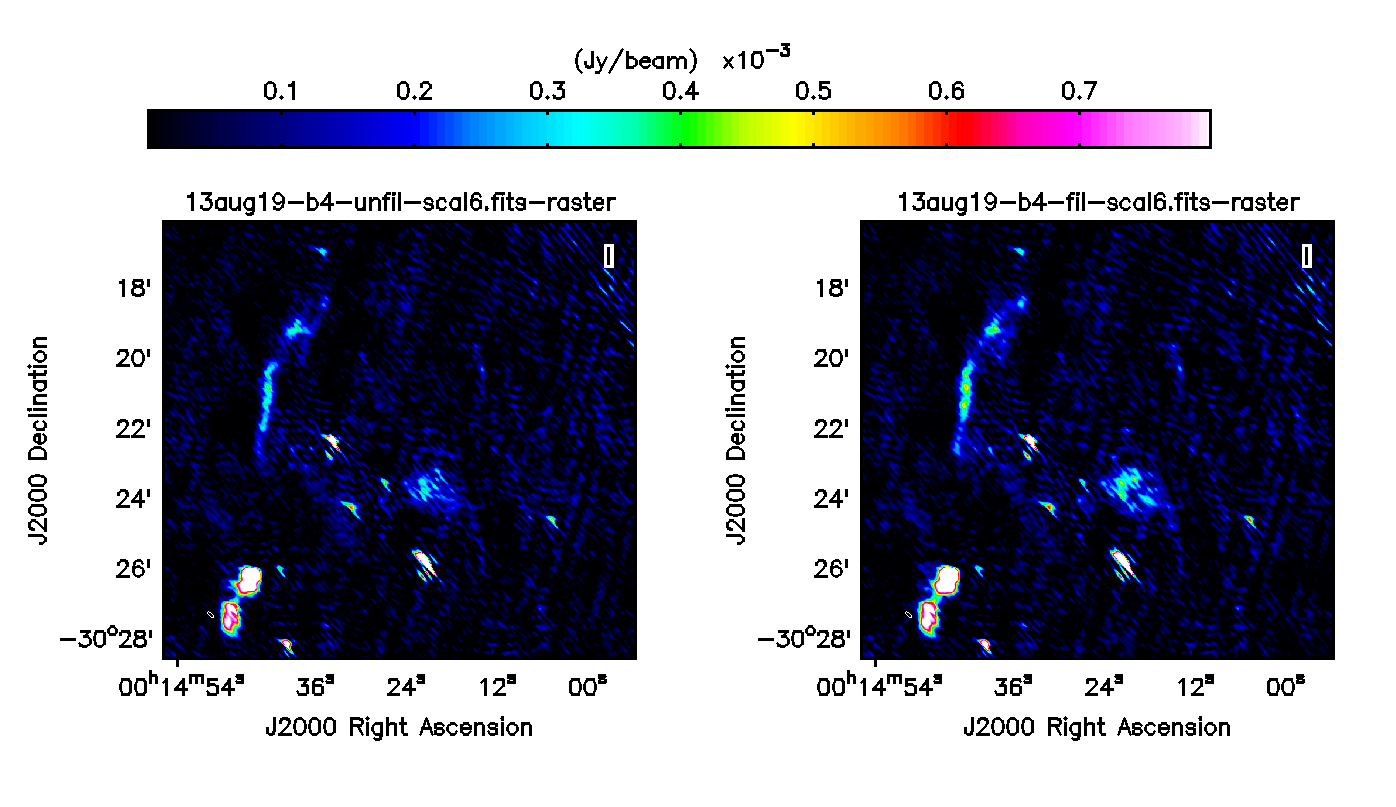}
 \caption{Band 4 images of A2744 field: unfiltered (left) and filtered (right). The rms in the unfiltered image is 44 $\mu$Jy beam$^{-1}$ and in the filtered image is 42 $\mu$Jy beam$^{-1}$. The beam size is $14.5''\times4.9''$, position angle $45.7^{\circ}$. [Sec.~\ref{flgimg}]
}
 \label{b4img}
 \end{center}
\end{figure}

\begin{figure}[h]
\begin{center}
      \includegraphics[trim={0cm 0cm 0cm 0cm}, clip,height=8cm]{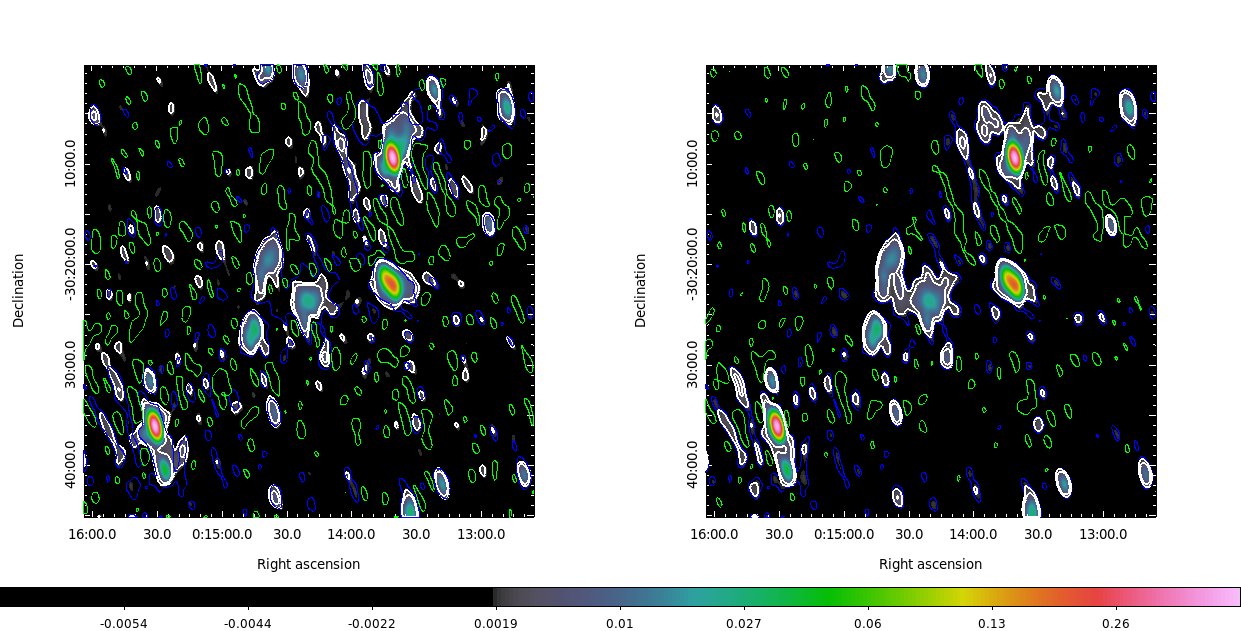}
 \caption{Image of A2744 made with use of baselines $<2$klambda at band 3 using unfiltered data (left) and filtered data (right). The contours are drawn at identical levels in the two panels. The green contours are at -0.0014, blue are at 0.0014 Jy beam$^{-1}$. This represents a level of $2\sigma$ where $\sigma$ is the rms in the filtered image. The white contours are at 0.0021 and 0.0042 Jy beam$^{-1}$ and represent $3$ and 6 $\sigma$ levels. The coordinates and the colour scale (Jy beam$^{-1}$) are matched in the two panels. [Sec.~\ref{flgimg}]
}
 \label{shortuvimg}
 \end{center}
\end{figure}

The images from the 1:2 tests carried out 
in bands 3 and 4 are shown in Figs.~\ref{b3img} and ~\ref{b4img}. At both the bands, there are minor but 
noticeable differences in the morphology of the extended emission. Qualitatively the appearance of the extended emission is smoother in the filtered image. 
We also show the images made from baselines shorter than $2$ kilolambda
 (band 3) in Fig.~\ref{shortuvimg}. The left and right panels show the images from 
 unfiltered and filtered data respectively, on an identical colour scale. The fidelity of the image from the filtered data is better, as indicated by the contours. The rms in the filtered and unfiltered images are 0.0007 Jy beam$^{-1}$ and $0.0009$ Jy beam$^{-1}$, respectively. The green contours are at $\pm 0.0014$ Jy beam$^{-1}$ 
 and the white contours are at 0.0021 and 0.0042 Jy beam$^{-1}$.

\section{Typical flagging percentages using the RFI counter} \label{counterpercent}
\begin{figure}[h]
\begin{center}
\includegraphics[trim={0cm 0cm 0cm 0cm}, clip,height=7.0cm]{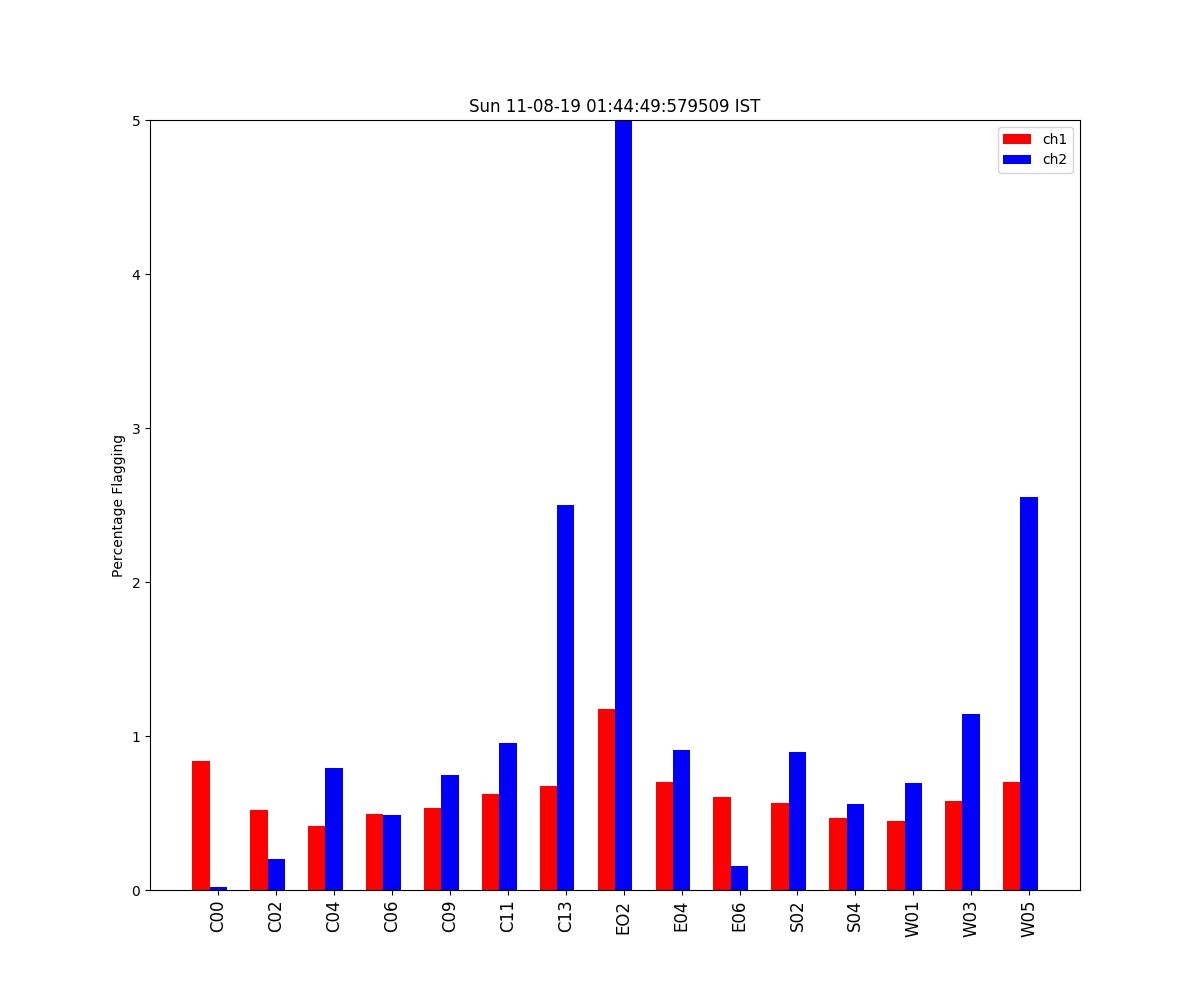}
        \includegraphics[trim={0cm 0cm 0cm 0cm}, clip,height=7.0cm]{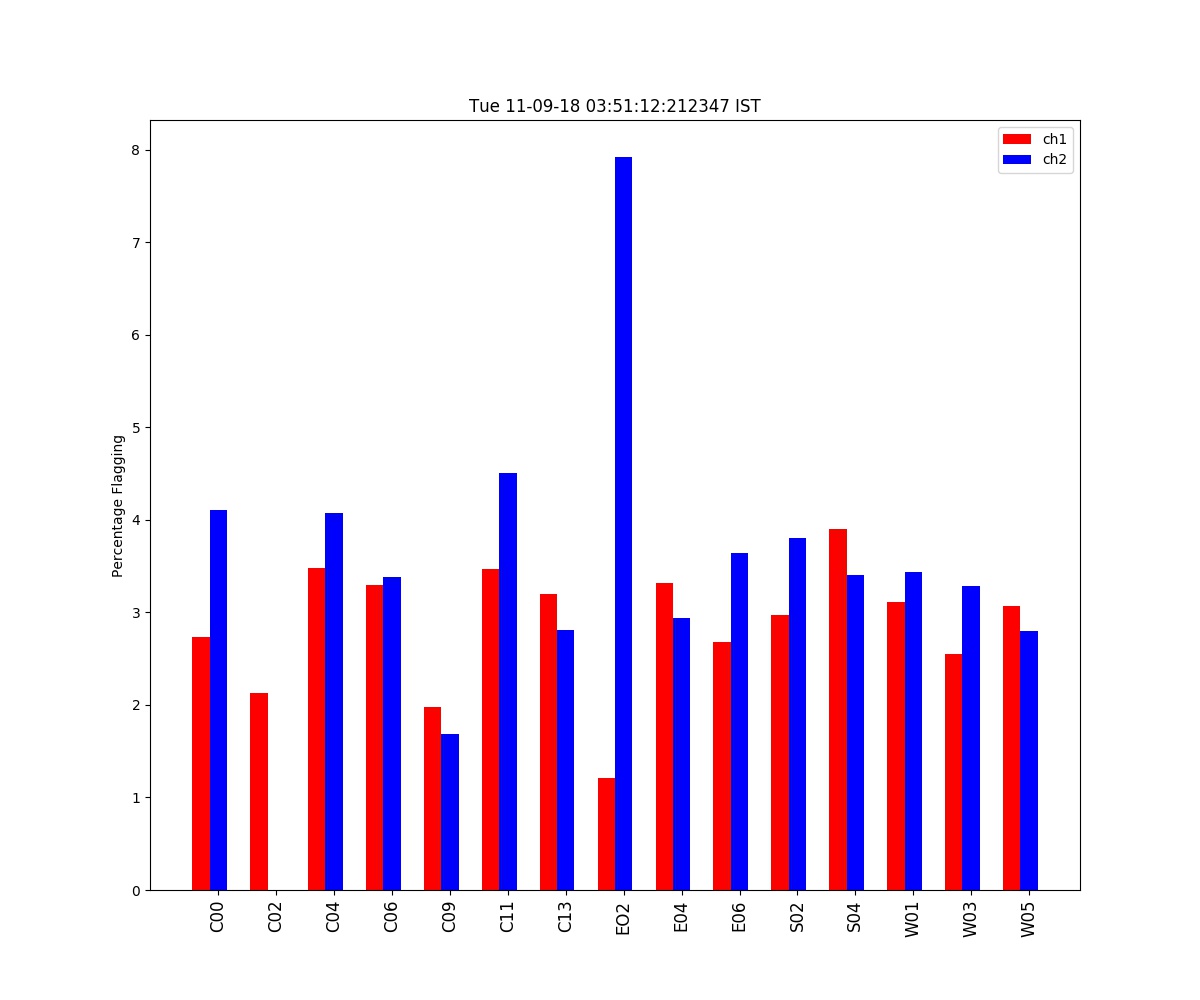}
        \includegraphics[trim={0cm 0cm 0cm 0cm}, clip,height=7.0cm]{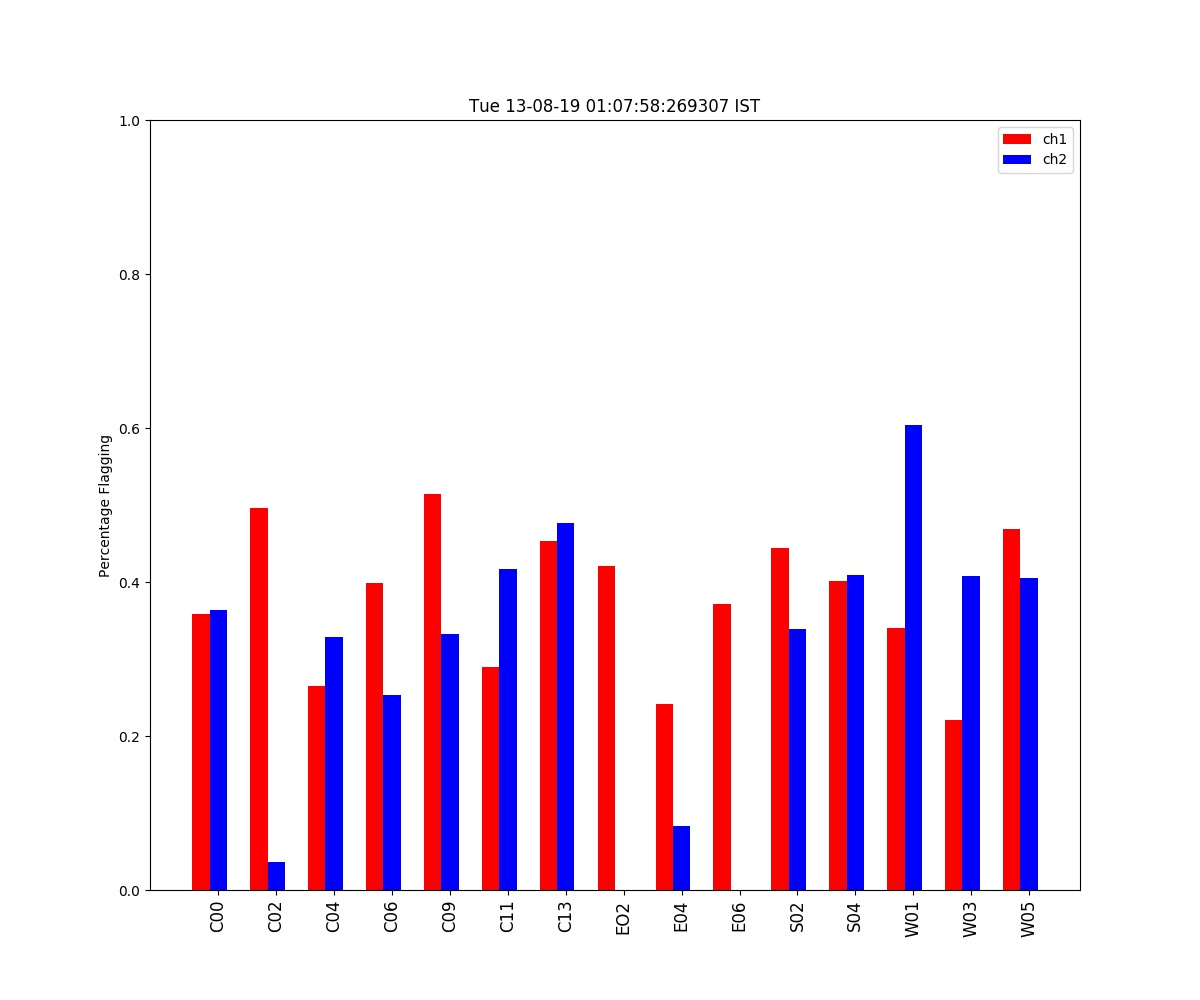}
         \includegraphics[trim={0cm 0cm 0cm 0cm}, clip,height=7.0cm]{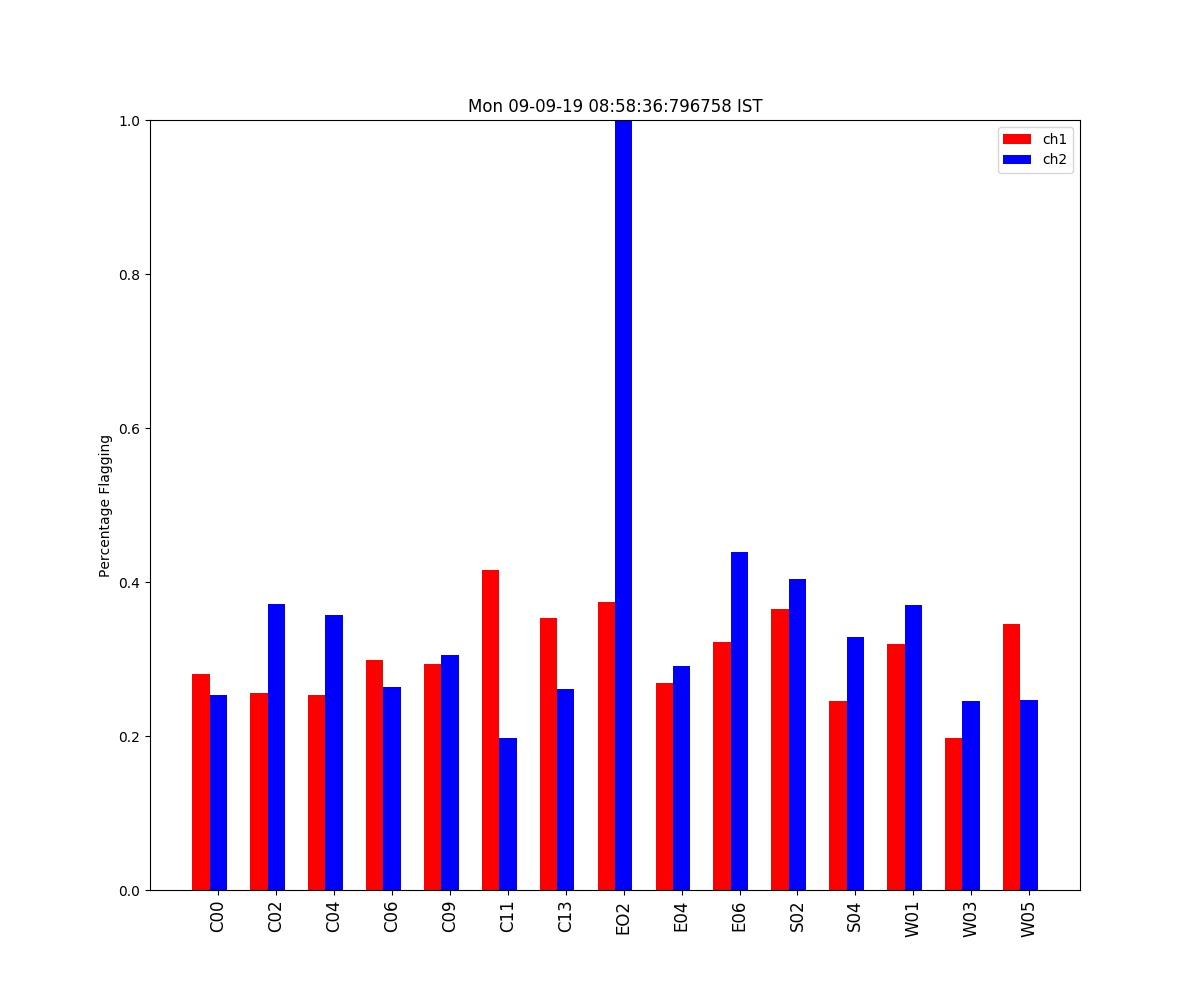}
 \caption{Band 2, 11 Sep. 2018, Band 3, 11 Aug. 2019, Band 4, 13 Aug. 2019 and Band 5, 09 Sep. 2019 counter histograms are shown. The two polarizations are shown in red (Channel 1) and blue (Channel 2) colours. The E02 antenna (channel 2) counts have a problem at all bands for one polarization. The typical flagging percentage is $<5\%$ for Band 2 (top-right), $<3\%$ for Band 3 (top-left), $<1\%$ for Band 4 (bottom-left) and $<0.5\%$ for Band 5 (bottom-right). [Sec.~\ref{counterpercent}]}
 \label{counterplots}
 \end{center}
\end{figure}

The RFI counter was used during the tests to keep track of the percentage of data affected by the filter. The flagging percentage values for the primary calibrator scan for each of the tests are shown in Fig.~\ref{counterplots}. 
 The typical flagging percentage is $<5\%$ for Band 2, $<3\%$ for Band 3, $<1\%$ for band 4 and $<0.5\%$ for Band 5.

\section{Results at Band 2: De-correlation}\label{band2section}
\begin{figure}[h]
\begin{center}
        \includegraphics[trim={2.5cm 1.5cm 1.5cm 2cm}, clip,height=10.0cm,,angle=90]{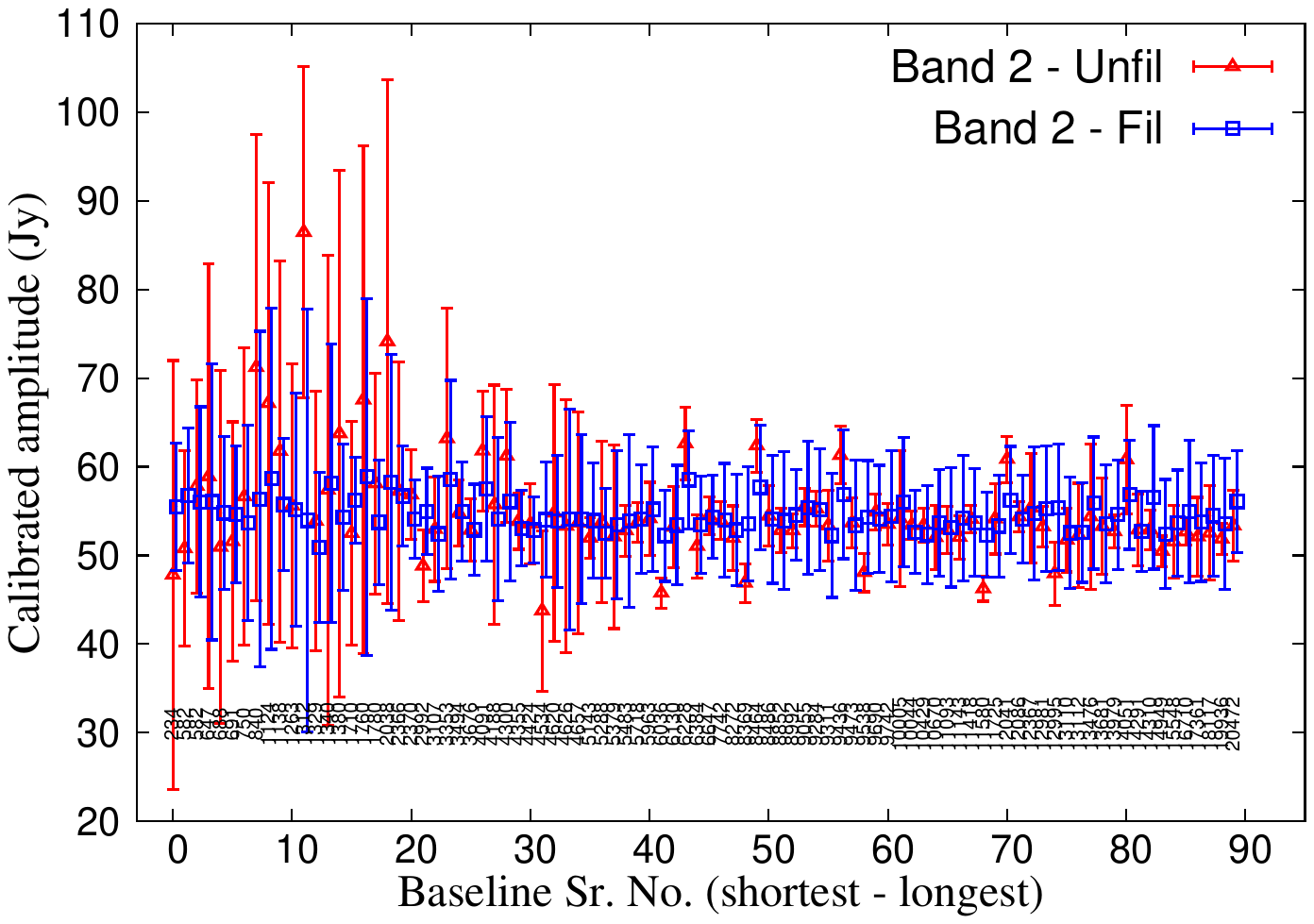}
 \caption{The band 2 calibrated flux density of a primary calibrator source for a single channel is plotted as a function of baseline numbers sorted according to their lengths. The integers in the plot 
 indicate the length of the baseline in metres. The unfiltered (red) and filtered (blue) points are offset on the x-axis for comparison. [Sec.~\ref{band2section}]}
 \label{b2plot}
 \end{center}
\end{figure}

\begin{figure}[h]
\begin{center}
        \includegraphics[trim={1cm 2.5cm 1cm 2cm}, clip,height=10.0cm]{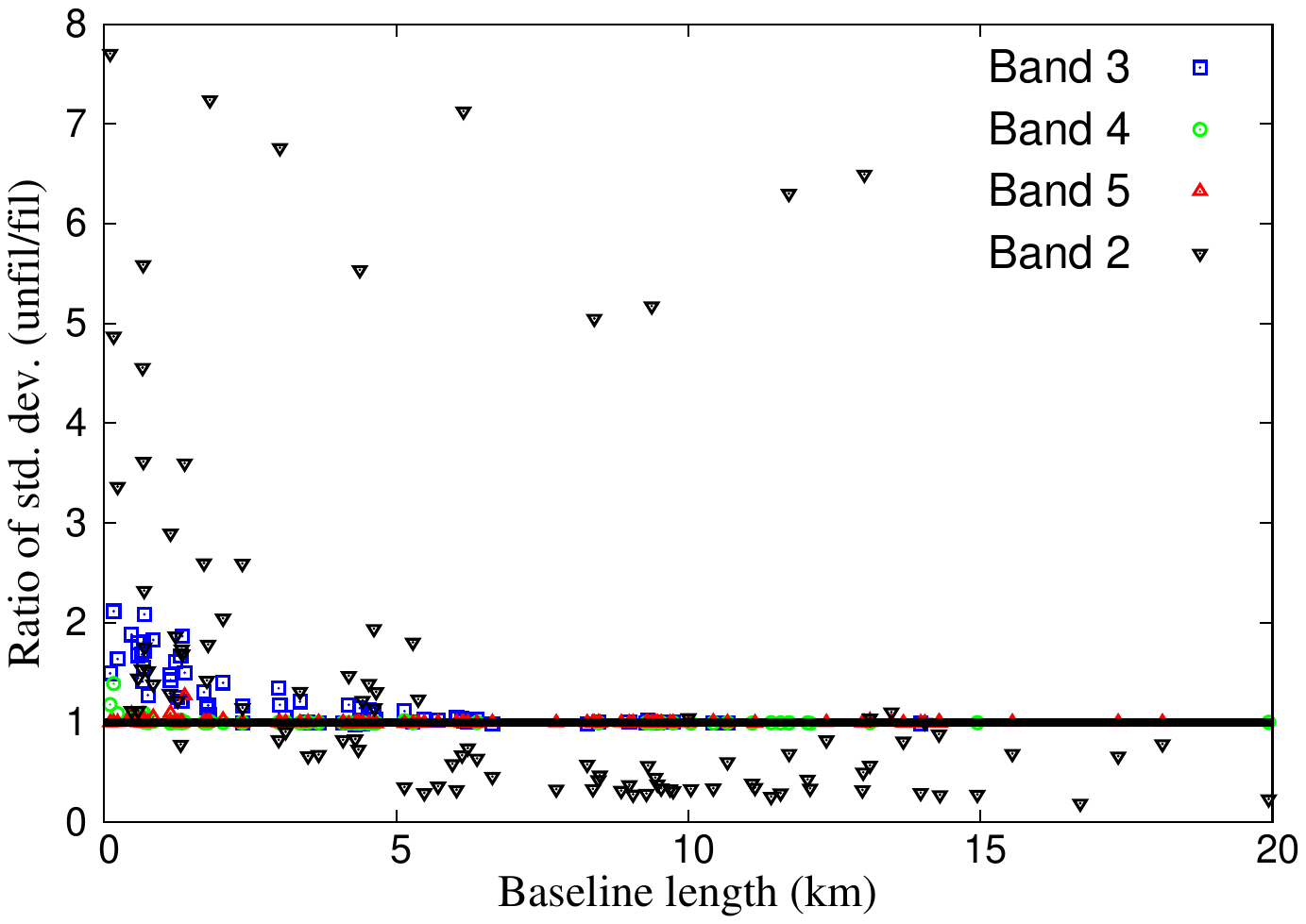}

 \caption{The ratio of standard deviations on the visibility amplitude (unfiltered/filtered) versus baseline length is shown for band 2 in black, and the remaining points are plotted from Fig.~\ref{b345plot}. [Sec.~\ref{band2section}]}
 \label{b2345plot}
 \end{center}
\end{figure}

At band 2, the data in the frequency range 0.19 - 0.24 GHz were analysed to avoid the regions in the spectrum 
that were affected by the notch filter and dominant narrow band RFI. The calibrated data towards the primary calibrator are shown in Fig.~\ref{b2plot}. The trend of reduced standard deviation in the data 
at short baselines seen at bands 3, 4 and 5 is also seen at band 2. However, there are long baselines that show an increase in the standard deviation after filtering. In Fig.~\ref{b2345plot}, the ratio of standard deviations for band 2 is plotted along with those for the bands 3, 4 and 5. The points with values below 1 indicate those baselines where the scatter has increased after filtering. The origin of this effect is likely the higher 
flagging percentage that introduces decorrelation in the data.


\section{Real-time BB RFI filter: Operational Aspects}

The RFI filter operates on 8-bit data, irrespective of the GWB processing bandwidth or the mode of operation. The performance of the RFI filter varies under different levels of input power (in the presence of strong NB RFI), and it is found that power equalization between 80-100 self counts for GWB is optimal. The power equalization needs to be done in spectral channels void of NB RFI and where the power level is stable and equal. Other recommended settings and the operating procedure for running RFI filtering in GWB is provided in \cite{gwbsop}, which is available with the GMRT control room. This filter is targeted only to remove BB RFI, and hence it may not affect the situation of NB RFI in the filtered data. The flagged fraction for each antenna for the entire observation or each scan be recorded. The details are available in the SOP. Currently, the command to record this count has to be explicitly entered in the observation command file.

\section{Summary and conclusions}
The real-time RFI filtering scheme has been implemented for the uGMRT. This scheme is particularly targeted for reducing the impact of broadband RFI on astronomical data. 
This system was tested by implementing a new testing scheme that allowed to recording 
the filtered and unfiltered data simultaneously. Tests were conducted in the upgrade slots and white slots available in the GMRT observing schedule. The conclusions from the tests and the current status of the system are as follows:
\begin{enumerate}
\item The engineering tests show an improvement in signal-to-noise ratio, cross-correlation function, and closure phase.
\item Beam former mode tests towards a standard calibrator show the removal of broadband features from the spectrum when the filter is used. Observation of the bright pulsar B0329+54 show an improvement by a factor of 4 in the signal-to-noise ratio. Improvements in single pulse detection and frequency-phasogram has also been found. 
\item Interferometric mode tests at bands 3, 4 and 5 show that the short baselines (central square) show significantly less number of outliers in amplitude and phases. The calibrator flux densities were recovered within a few per cent of the expected value, and no systematic was noticed between filtered and unfiltered data. 
\item On analysing the filtered and unfiltered data with an identical analysis pipeline, it was found that the flagging percentage was reduced in filtered data as compared to unfiltered data by a few to ten per cent at bands 3 and 4. 
\item Images produced using the filter had improvement in RMS and had better image fidelity. \item An RFI counter also has been implemented, and it shows that  $5\%, 3\%$, $1\%$ and $0.5\%$ of data samples are replaced by digital noise at bands 2, 3, 4 and 5, respectively. 
\item Decorrelation is found to be introduced in filtered data, especially at band ,2, where the percentage of the RFI is the highest. At bands 3 and 4, the effect of decorrelation is less than $2.5\%$. We are carrying out further tests to understand the decorrelation.
\item The real-time RFI filtering with $3\sigma$ threshold and digital noise replacement has been tested and is released on a shared risk basis from Cycle 38 onwards. A SOP has been written for the control room. To use the RFI counter, additional commands need to be added to the command file, which can be obtained by writing to us (onlinerfi@ncra.tifr.res.in). A short reference document for GTAC users has also been written. 
\item  Further tests to recommend optimal thresholds for different bands are being carried out. Testing for spectral line observations is also being carried out.
\end{enumerate}

\section*{Acknowledgements}
The authors would like to thank the GMRT Operations, GMRT Backend and Control room staff for their help in carrying out the observations. The engineering team would like to thank all those who worked towards the development and testing of the real-time RFI excision system (in particular Kishor Naik, Swapnil Nalawade and Shruti Bhatporia). We would like to thank Kshitija Pawar for helping with the formatting of this report. We would like to thank Jayanta and Bhaswati for the planning, observational support and data analysis
for the pulsar observations described in this report. We would like to thank Prof. Yashwant Gupta for his guidance and support, and all the NCRA astronomers for their valuable inputs at each stage of development and testing. We thank Prof. Govind Swarup and Prof. Willem Baan for providing guidance and encouragement in this effort.

\bibliography{rfi_report}   
\bibliographystyle{ieeetr}      

\begin{table}[h]
\centering
\caption{Summary of calibrator flux densities in filtered and unfiltered data. [Sec.~\ref{calflux}]}\label{calsrcflux}
\vspace*{0.2cm}
 \begin{tabular}{cccccccc}
 \hline
 \hline
  Date           & Band & Mode              & Source  & Un-filtered & Filtered & Fractional\\
                 &      &                   &         & Jy         & Jy & change ($\%$)\\

  \hline
  14 Jun. 2018   &  4  & 1:2               & 3C48     & $34.3\pm0.2$&$33.6\pm0.1$ &2\\
    \hline

                   \hline

  11 Aug. 2019   & 3   & 1:2               & 3C48     &$38.2\pm3.6$ &$38.1\pm2.9$ &0.3\\
                 &     &                   & 0116-208 &$10.0\pm2.0$ &$9.7\pm1.3$ &3.0\\
  \hline

  13 Aug. 2019   & 4   & 1:2               & 3C48     &$30.6\pm3.3$ &$30.6\pm3.2$ &0\\
                 &     &                   & 0116-208 &$7.8\pm0.9$ &$7.9\pm0.9$ &1.2\\
 09 Sept. 2019   & 5   & 1:2               & 3C147    &$22.6\pm3.0$&$22.5\pm3.0$ & 0.4\\
   \hline
   \hline
 \end{tabular}
\end{table}

\begin{table}[]
\centering
\caption{Summary of flagging on target source.[Sec.~\ref{flgimg}]}\label{flagstat}
\vspace*{0.2cm}
 \begin{tabular}{cccccccc}
 \hline
 \hline
  Band  & Unfiltered  & Filtered & Unfiltered & Filtered \\
        &Flagged ($\%$)&Flagged ($\%$) &Flagged ($\%$)& Flagged ($\%$) \\
       &   $<2klambda$ &  $<2klambda$ &all uvrange &all uvrange \\
       \hline
   3   &40.2 & 36.1 & 53.7 & 53.7  \\          
  \hline
   4   &15.0 & 9.1& 33.6 & 32.7 \\
   \hline
 \end{tabular}
\end{table}

\end{document}